\documentclass[12pt,notitlepage]{article}

\usepackage{amsmath}
\usepackage{graphicx}
\usepackage[english]{babel}
\usepackage[utf8]{inputenc}
\usepackage[T1]{fontenc}
\usepackage{setspace}
\usepackage[english]{varioref}
\usepackage{textcomp}
\usepackage{endnotes}
\usepackage{amsthm}
\usepackage{rotating}
\usepackage{stmaryrd}
\usepackage{amssymb}
\usepackage{xcolor}
\usepackage[affil-it]{authblk}
\usepackage[caption=false]{subfig}
\usepackage{pifont}
\usepackage{color}
\usepackage{float} % for [H] float positioning
\usepackage{breakcites} % to avoid citations running into the right margin
\usepackage{microtype} % to avoid citations running into the right margin
\usepackage{xfrac}
\usepackage[width=.8\textwidth]{caption}

\usepackage{siunitx}             % Proper formating for SI units
\usepackage[version=4]{mhchem}	 % Chemical formulae
\usepackage{enumitem}           % Formatting of itemize and enumerate styles

\usepackage{hyperref}
\usepackage{color}
\definecolor{linkcol}{rgb}{0,0,0.4} 
\definecolor{citecol}{rgb}{0.5,0,0} 

\DeclareSIUnit\bar{bar}

\newcommand{\isotope}[2]{$^{#2}{\rm #1}$}

\begin{document}
\thispagestyle{empty}
%%%%%%%%%%%%%%%%%%%%%%%%%%%%%%%%%%%%%%%%%%%%%%%%%%%%%%%%%%%
%
%	Coverpage
%
%%%%%%%%%%%%%%%%%%%%%%%%%%%%%%%%%%%%%%%%%%%%%%%%%%%%%%%%%%%
\title{\bf The Project 8 Neutrino Mass Experiment}
\author[1]{A.~Ashtari~Esfahani}
\author[2]{S.~B\"oser}
\author[3]{N.~Buzinsky}
\author[4]{M.~C.~Carmona-Benitez}
\author[1]{C.~Claessens}
\author[4]{L.~de~Viveiros}
\author[1]{P.~J.~Doe}
\author[1]{S.~Enomoto}
\author[2]{M.~Fertl}
\author[3]{J.~A.~Formaggio\thanks{josephf@mit.edu}}
\author[5]{J.~K.~Gaison}
\author[5]{M.~Grando}
\author[6]{K.~M.~Heeger}
\author[5]{X.~Huyan\thanks{Present Address: LeoLabs, Inc., Menlo Park, CA 94025, USA}}
\author[5]{A.~M.~Jones}
\author[7]{K.~Kazkaz}
\author[3]{M.~Li}
\author[2]{A.~Lindman}
\author[2]{C.~Matth\'e}
\author[8]{R.~Mohiuddin}
\author[8]{B.~Monreal}
\author[4]{R.~Mueller}
\author[6]{J.~A.~Nikkel}
\author[1]{E.~Novitski}
\author[5]{N.~S.~Oblath}
\author[3]{J.~I.~Pe\~na}
\author[9]{W.~Pettus\thanks{pettus@indiana.edu}}
\author[2]{R.~Reimann}
\author[1]{R.~G.~H.~Robertson\thanks{rghr@uw.edu}}
\author[1]{G.~Rybka}
\author[6]{L.~Salda\~na}
\author[5]{M.~Schram}
\author[6]{P.~L.~Slocum}
\author[3]{J.~Stachurska}
\author[8]{Y.-H.~Sun}
\author[6]{P.~T.~Surukuchi}
\author[5]{J.~R.~Tedeschi}
\author[6]{A.~B.~Telles}
\author[2]{F.~Thomas}
\author[5]{M.~Thomas\thanks{Present Address: Booz Allen Hamilton, San Antonio, Texas, 78226, USA}}
\author[2]{L.~A.~Thorne}
\author[10]{T.~Th\"ummler}
\author[3]{W.~Van~De~Pontseele}
\author[1,5]{B.~A.~VanDevender\thanks{brent.vandevender@pnnl.gov}}
\author[6]{T.~E.~Weiss}
\author[4]{T.~Wendler}
\author[4]{A.~Ziegler}

\affil[1]{Center for Experimental Nuclear Physics and Astrophysics and Department of Physics, University of Washington, Seattle, WA 98195, USA}
\affil[2]{Institute for Physics, Johannes-Gutenberg University Mainz, 55128 Mainz, Germany}
\affil[3]{Laboratory for Nuclear Science, Massachusetts Institute of Technology, Cambridge, MA 02139, USA}
\affil[4]{Department of Physics, Pennsylvania State University, University Park, PA 16802, USA}
\affil[5]{Pacific Northwest National Laboratory, Richland, WA 99354, USA}
\affil[6]{Wright Laboratory, Department of Physics, Yale University, New Haven, CT 06520, USA}
\affil[7]{Lawrence Livermore National Laboratory, Livermore, CA 94550, USA}
\affil[8]{Department of Physics, Case Western Reserve University, Cleveland, OH 44106, USA}
\affil[9]{Department of Physics, Indiana University, Bloomington, IN, 47405, USA}
\affil[10]{Institute of Astroparticle Physics, Karlsruhe Institute of Technology, 76021 Karlsruhe, Germany}

\date{\today}
\maketitle 

\begin{abstract}
Measurements of the $\beta^-$ spectrum of tritium give the most precise direct limits on neutrino mass. Project 8 will investigate neutrino mass using Cyclotron Radiation Emission Spectroscopy (CRES) with an atomic tritium source. CRES is a new experimental technique that has the potential to surmount the systematic and statistical limitations of current-generation direct measurement methods. Atomic tritium avoids an irreducible systematic uncertainty associated with the final states populated by the decay of molecular tritium. Project 8 will proceed in a phased approach toward a goal of 40 meV/c$^2$ neutrino-mass sensitivity. \\ \\
\begin{center}
    Submitted to the Proceedings of the US Community Study\\
    on the Future of Particle Physics (Snowmass 2021).\\
\end{center}
\end{abstract}

\newpage
\tableofcontents

\section{Executive Summary}\label{sec:Summary}

Project 8 is a novel experiment designed to measure the absolute mass scale of the neutrino.    Though the existence of neutrino mass is now firmly established experimentally, the mass scale itself is still unknown, and remains an outstanding question in the field of experimental neutrino physics.  Knowledge of the neutrino mass scale has far reaching implications in nuclear physics, particle physics, and cosmology.  Pursuit of this question is a matter of high priority to the scientific community.

The Project 8 experiment employs a new technique by which the energy spectrum of low energy electrons can be extracted.  The technique relies on the detection and measurement of coherent radiation created from the cyclotron motion of electrons in a strong magnetic field.  Detection and measurement of the coherent radiation emitted is tantamount to measuring the kinetic energy of the electron.  As the technique inherently involves the measurement of a frequency without destructively interacting with the electron, it can, in principle, achieve a high degree of precision and accuracy.  One immediate application of this technique is in the measurement of the endpoint spectrum from tritium beta decay, which is directly sensitive to the absolute mass scale of neutrinos.  In addition to its primary mission to measure the absolute neutrino mass scale, the Project 8 experiment is also sensitive to a number of other processes, including the neutrino mass ordering, sterile neutrinos, and the over-abundance of relic neutrinos.

Over the past five years, the Project 8 collaboration has been able to execute a proof-of-principle measurement using cyclotron radiation emission spectroscopy (CRES) on both radioactive krypton (\isotope{Kr}{83m}) and molecular tritium (\isotope{H_2}{3}).  These measurements demonstrate both the  high accuracy of the frequency technique as well as its inherent low background.  These measurements have  yielded a detailed understanding of the underlying physics of CRES, allowing for an accurate modeling of these processes to develop solid projections.  Finally, the measurement program has yielded the first neutrino mass limit using the CRES technique.

Over the next few years, the Project 8 collaboration plans to expand the experiment to enlarge the sensitive volume of the experiment, while still retaining high accuracy and low background.  The collaboration is currently evaluating two possible technical approaches: a free space CRES demonstrator, which uses rings of antennas to reconstruct electrons, or a mode-filtered resonant cavity.  The collaboration is also evaluating moving the operating frequency from 26 GHz to 1 GHz.  The final technology design choice will be determined based on comparisons on achievable detector efficiency, energy resolution and scalability.  In parallel, the collaboration is also pursuing the development of a cold (few mK) atomic tritium source.  Switching from molecular to atomic tritium will remove one of the remaining systematic uncertainties in tritium beta decay due to molecular tritium final states.  Much work has already taken place in developing a high luminosity atomic (hydrogen) source. The design work on the cooling and transport of an atomic beam continues.  The atomic source would eventually be stored in an atomic magneto-gravitational trap.  Both Ioffe or Halbach array magnetic traps are currently being evaluated.  

The Project 8 collaboration is undertaking such a staged approach to measuring the neutrino mass scale, so that it can achieve its final goal of measuring the neutrino mass scale down to the inverted hierarchy scale of 40 meV/c$^2$.  Achieving this goal will have significant impacts on the fields of nuclear physics, particle physics, and cosmology.
\section{Scientific Motivation}

{\em The aim of the Project 8 neutrino mass experiment is to provide a direct and model-independent laboratory constraint on the neutrino mass by measuring the kinetic energy spectrum of tritium decay electrons. In addition to its primary objective, Project 8 is also sensitive to a myriad of other processes, including the neutrino mass ordering, sterile neutrinos, and over-abundance of relic neutrinos.}
%The motivation behind these measurements is described in more detail below. Pranava: I think this is redundant.

\subsection{Neutrino Mass}\label{sec:neutrino_mass}

Experiments demonstrating non-zero neutrino mass following from flavor oscillation phenomena are now legion~\cite{Fukuda:1998fk, Ahmad:2001fk, Ahmad:2002fk, Eguchi:2003kx, Aharmim:2008kc, Abe:2011ys, An:2012ve, Ahn:2012ly}.  However, oscillation phenomena depend only on the differences of the squares of neutrino mass eigenvalues $\Delta m_{ij}^2 \equiv m_i^2 - m_j^2$; the absolute mass scale does not enter this description.  Furthermore, two independent $\Delta m_{ij}^2$ magnitudes have been measured~\cite{Fukuda:1998fk, Ahmad:2001fk}, but the sign is determined only for one of them. This leads to ambiguity in the ordering of mass eigenstates, with a ``normal'' and an alternative ``inverted'' possibility.  Oscillation phenomena do imply a lower limit on the mass scale $m_\beta$ determined in laboratory experiments since the lightest possible eigenvalue is zero:  $m_{\beta} > 0.048$\,eV for inverted ordering, and  $m_{\beta} > 0.0085$\,eV for normal ordering (95\% confidence level)~\cite{Zyla:2020aa}.

\begin{figure}
    \centering
    \includegraphics[width=0.45\textwidth]{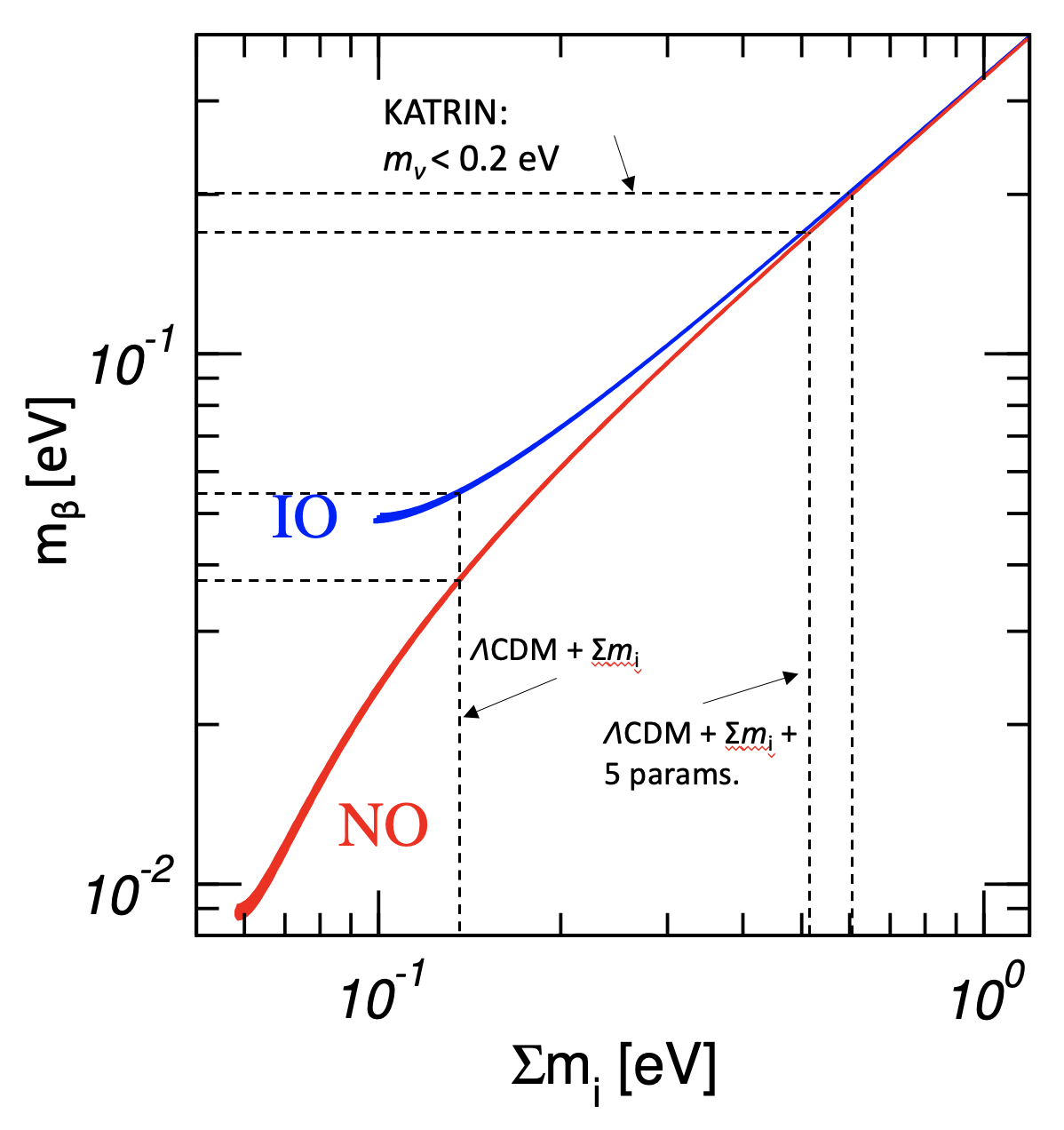}
    \caption{The relationship between the neutrino mass observable in tritium endpoint experiments ($m_\beta$) and the observable from cosmology ($\Sigma m_i$).  Figure reproduced from~\cite{Esteban:2019aa}, with annotations added.}
    \label{fig:beta_vs_cosmo}
\end{figure}

Only a handful of methods can probe the absolute neutrino mass scale. Cosmological observation and data interpretation within cosmological models provide an indirect method for probing the neutrino mass.  Hot Big Bang scenarios generally feature a cosmic neutrino background. The time at which those neutrinos become non-relativistic depends on their mass and leaves a distinct imprint on the cosmic microwave background (CMB) anisotropy. At late epoch, neutrinos become non-relativistic, which tends to suppress structure on smaller scales in the matter power spectrum as seen, for example, in baryon acoustic oscillations (BAOs). The Planck satellite limit with these combined observations is $\Sigma m_i \leq \SI{0.12}{\electronvolt}$~\cite{Aghanim:2020aa}.  That is more restrictive than KATRIN's projected limit (see Figure~\ref{fig:beta_vs_cosmo} for the correspondence between $\Sigma m_i$ and $m_\beta$).
However, if cosmology turns out to require more free parameters, the limit is significantly relaxed.  One compelling extension includes six parameters beyond the usual six of $\Lambda$CDM: the sum of neutrino masses, the effective number of flavors, and four others~\cite{Di-Valentino:2015aa}. In that context the limit is $\Sigma m_i <$ \SI{0.52}{\electronvolt}~\cite{Tanabashi:2018aa}, comparable to KATRIN's projected limit.  It is also the case that fixing any parameter, such as by measuring neutrino mass independently in the laboratory, would improve sensitivity to the remaining parameters that can only be extracted from cosmological data.  This scenario motivates direct measurements like Project 8, even in the coming age of precision cosmology, and especially if tensions like the current Hubble crisis~\cite{Di-Valentino:2020aa} persist or new ones arise.

In particle physics, a determination of the scale and nature of neutrino masses could provide valuable insight into the mechanism that generates masses.  Given the extreme disparity in mass scales, it seems unlikely that neutrinos acquire mass by the same Higgs coupling as the other fundamental fermions.  It is not known whether neutrinos are Dirac or Majorana in nature. The direct mass measurement method is independent of this nature.  Neutrinos must be Majorana for neutrinoless double-beta decay to occur, and direct measurements can thus provide guidance on required sensitivity in searches for this process.  A number of theories beyond the Standard Model make predictions of the scale and ordering of the neutrinos, so direct neutrino mass measurements will play an important role in the resolution of this fundamental question. 

\subsection{Neutrino Mass Ordering}

With sufficient sensitivity, it is possible to utilize the tritium $\beta$-decay technique's sensitivity to spectral distortions to extract not just the neutrino mass scale, but also the mass ordering (inverted or normal). For neutrino mass scale sensitivity below 50 meV, it is possible to discern the ``kink" that develops due to the distinct mass eigenstates, using existing constraints on the neutrino mixing parameters from reactor experiments.  A detailed study of the potential sensitivity of a high resolution direct neutrino mass experiment can be found in~\cite{AshtariEsfahani:2020bfp}. When sensitivity to the lightest mass is better than $m_\beta \le 0.03$\,eV/c$^2$, it is nearly always possible to resolve the mass ordering.

\begin{figure}
    \includegraphics[width=0.5\textwidth]{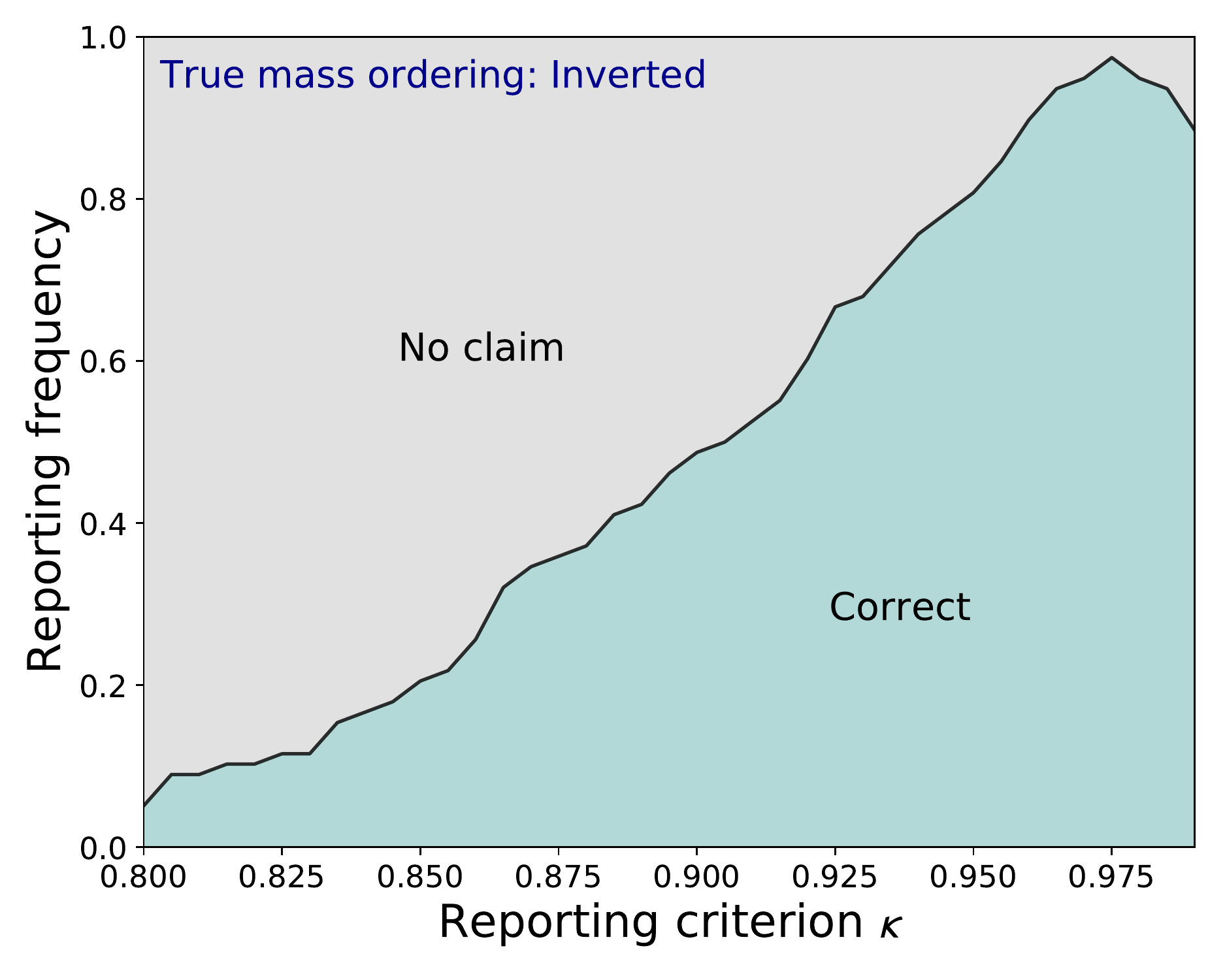}
    \includegraphics[width=0.5\textwidth]{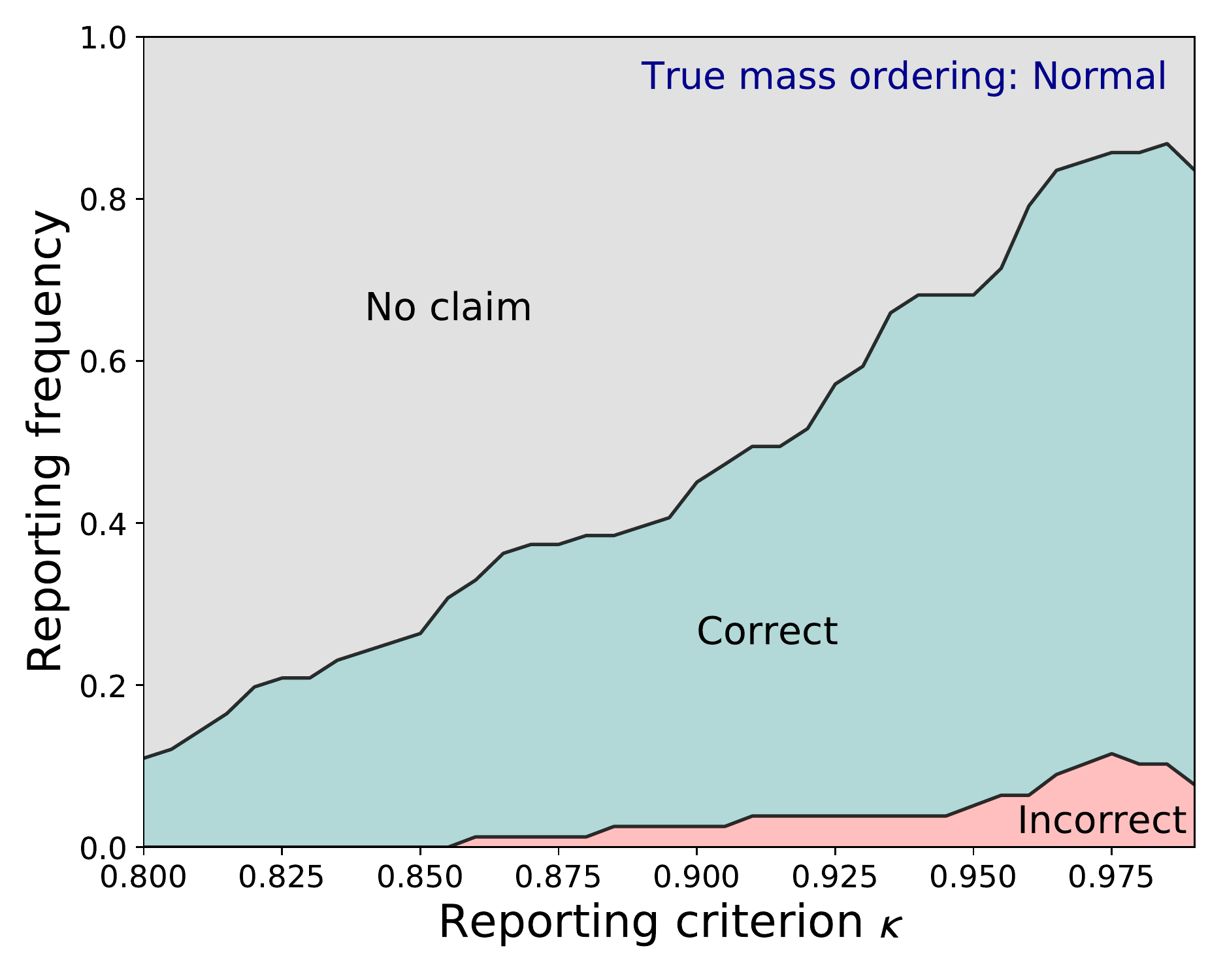}
    \caption{Mass ordering reporting frequencies for $\Delta t=2$\,yrs as a function of the credibility of the $\eta$ interval ($\eta$ is the fractional contribution of the lighter mass term to the spectral shape.). See~\cite{AshtariEsfahani:2020bfp} for additional details.}
\label{fig:ordering}
\end{figure}

\subsection{Sterile Neutrinos}

By relaxing the constraint on the observed mass splittings and mixing parameters, it is possible to utilize access to the spectral energy distribution from beta decay to search for additional kinematic distortions.  Evidence of such distortions would be particularly sensitive to additional neutrino mass eigenstates, such as sterile neutrinos. As a differential spectroscopy method, CRES allows Project 8 to make simultaneous searches for both active and sterile masses. Furthermore, the benefits of the CRES technique for neutrino mass measurement---namely low backgrounds, good resolution, and high event rates---also apply to a search for sterile neutrinos. Hence, a superior sensitivity to the direct neutrino mass also provides superior sensitivity to a sterile neutrino.

\begin{figure}[htb]
    \centering
    \includegraphics[width=0.85\textwidth]{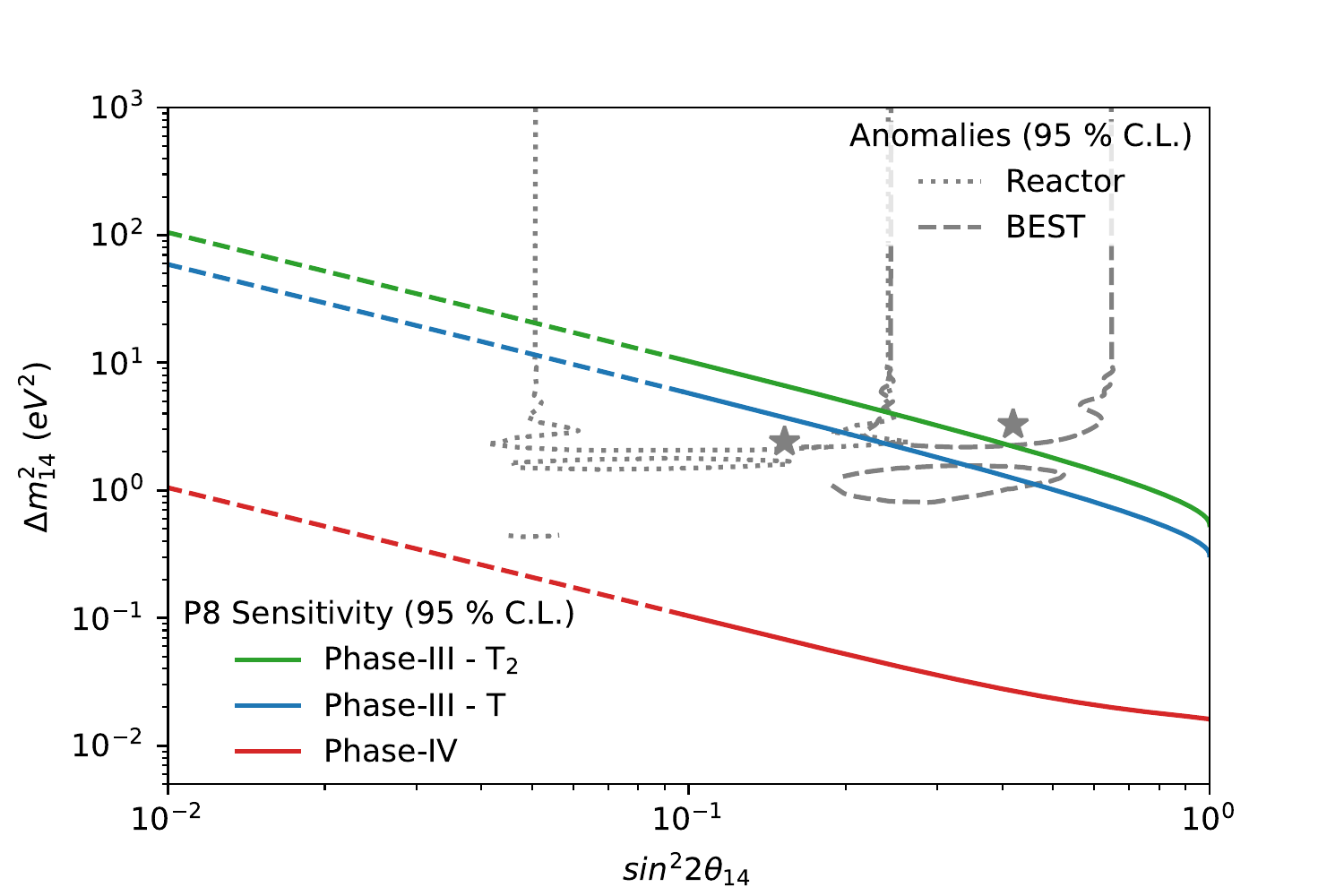}
    \caption{Sensitivity of the upcoming phases of Project 8 experiment to light sterile neutrinos in the 3+1 framework. All curves, including reactor and BEST suggested parameters, are shown at 95 \% C.L. 
    Phase-IV of Project 8 aims to completely cover the reactor and BEST gallium anomalies at high significance. }
    \label{fig:steriles}
\end{figure}

The sensitivity of Project 8 to sterile neutrinos is determined using the analytical neutrino mass sensitivity method as suggested in Ref.~\cite{direct} and described in Section.~\ref{subsec:Sensitivity}. Figure~\ref{fig:steriles} shows the upcoming phases of Project 8 to be capable of a competitive sterile neutrino search over several orders of magnitude in $\Delta m^2_{14}$. 
The sensitivity is statistically limited and includes the current best knowledge of the systematics from energy resolution arising from thermal Doppler broadening, frequency to energy conversion, and variations in magnetic field. Control of systematic uncertainties leading to sensitivities in $\sin^2(2\theta) > 0.1 $ (shown as a solid line) is expected to be relatively straightforward. Further careful systematic control could enable experimental sensitivities down to $\sin^2(2\theta) \sim 0.01$ (shown as dashed lines). In the near future, with Phase-III, Project 8 aims to reach down to $\Delta m^2_{14}\sim$ eV$^2$ (C.L. 95\%) and cover major portions of the reactor~\cite{Abazajian:2012ys} and BEST~\cite{Barinov:2021asz} gallium anomaly suggested parameter spaces including the BEST best-fit point.
The experiment's sensitivity to higher $\Delta m^2_{14}$ is primarily limited by the efficiency of the cyclotron frequency detection method for lower energy electrons since a higher value of $m_4$ would manifest as a kink at a lower energy in the $\beta$-decay spectrum.
In generating sensitivity curves in Figure~\ref{fig:steriles}, the efficiency was assumed to be well-understood for energies tens of eV below the endpoint.
This is a fair assumption based on Project 8's ability to quantify the efficiency of the complex Phase-II detector over 2.5 keV below the endpoint.
The two different detection methods being investigated for upcoming phases are yet to demonstrate the control of efficiency over the wide energy range, but are expected to have lower complexity in efficiency than in Phase II.
Project 8 through the differential $\beta$-decay spectrum measurement using CRES thus provides a promising avenue to search for light sterile neutrinos in the near future.

\subsection{Relic Neutrinos}

Sensitivity to the relic density of cosmological neutrinos has always been considered a byproduct of every direct neutrino mass experiment which employs beta decay or electron capture.  The detection mechanism is through neutrino capture~\cite{Weinberg1962}:
\begin{equation}\label{eqn:neutrinoCapture}
\nu_{e} + ^3{\rm H} \to \, ^3{\rm He}^+ + e^-.
\end{equation}

Despite its relatively clean signature of a near mono-energetic peak located at an energy $2 m_\beta$ above the observed endpoint, it is hampered by the extremely low yields expected assuming standard relic neutrino density of 56 cm$^{-3}$ per neutrino flavor and chirality, approximately 10 events/year per 100 grams of tritium~\cite{Betts:2013uya}.  Recently, the KATRIN collaboration released a limit on the relic neutrino density of  of $9.7 \times 10^{10}$ cm$^{-3}$ at 90\% confidence level~\cite{KATRIN:2022kkv}.  Project 8 is expected to also have sensitivity to a relic neutrino overabundance.  Access to the differential energy spectrum from tritium beta decay should yield lower backgrounds and hence enhanced sensitivity to the process.
\section{Tritium Beta Decay and the CRES Technique}\label{sec:beta_decay}

{\em A direct and model-independent laboratory constraint on the neutrino mass can be derived from the kinematics of beta decay or electron capture. The most promising place to look for the absolute scale of neutrino mass is in the kinematics of tritium beta decay.}

\subsection{The Beta Decay Spectrum}

In beta decay, the energy available from the nuclear mass difference is carried away by the electron, the neutrino, and the progeny nucleus.  The three particles share the energy in a statistical way, determined quantum mechanically by the available phase space for each.  If the neutrino has rest mass, that small amount of energy alters the electron spectrum near its endpoint where it would otherwise have taken all the energy.  

The relative influence of neutrino mass on the spectrum compared to the available energy is maximized by choosing isotopes with the smallest endpoint energies (Q-values).  As such, isotopes with the lowest endpoint energy are typically the target for direct neutrino mass experiments.  These include \isotope{H}{3}, \isotope{Ho}{163}, \isotope{Re}{187}, \isotope{Cs}{135}, and \isotope{In}{115}~\cite{direct}.  The most sensitive limits on the neutrino mass scale have come from the beta decay of tritium~\cite{Otten:2008zz}:

\begin{equation}\label{eqn:betaDecay}
^3{\rm H} \to \, ^3{\rm He}^+ + e^- + \overline{\nu}_{e}.
\end{equation}

The electron energy spectrum of $\beta$-decay for a neutrino with
component masses $m_1$, $m_2$, and $m_3$ is the incoherent sum of the contributions from each mass eigenstate:
\begin{eqnarray}
\frac{d\Gamma}{dE} &=&  \frac{G_F^2|V_{ud}|^2}{2\pi^3}(G_V^2+3G_A^2)F(Z,\beta) \beta (E+m_e)^2(E_0-E) \nonumber \\&& \times \sum_{i=1,3}|U_{ei}|^2\left[(E_0-E)^2-m_i^2\right]^\frac{1}{2} \Theta (E_0-E-m_i), \label{eq:mother}
\end{eqnarray}
\noindent where $G_F$ is the Fermi coupling constant, $V_{ud}$ is an element of the CKM matrix \cite{Zyla:2020zbs}, and $E$ ($\beta$) denotes the electron's kinetic energy (velocity).
$E_0$, the `endpoint energy,' corresponds to the maximum kinetic energy in the absence of neutrino mass.
$F(Z,\beta)$
is the Fermi function, taking into account the Coulomb interaction
of the outgoing electron in the final state.
$\Theta (E_0-E-m_i)$ is the step function that ensures energy conservation. The vector and axial-vector matrix elements are $G_V=1$ and $G_A=-1.2646(35)$ for tritium, respectively \cite{Akulov:2005umb}.

In the approximation that the energy resolution exceeds the level splittings of the neutrino mass states, the sum across different mass eigenstates can be reduced to a single term, $m_{\beta}$, which represents the electron-weighted neutrino mass.
This is constructed as the incoherent sum of the neutrino mass states, weighted by the PMNS mixing matrix elements $U_{ei}$:

\begin{equation}
    m_{\beta}^2 = \sum_{i=1}^3 \left| U_{ei} \right|^2 m_i^2.
\label{eqn:m_beta}
\end{equation}

\noindent
We usually refer to $m_\beta$ as defined by Equation~\ref{eqn:m_beta} as ``the neutrino mass'' for brevity, even though the particle involved is actually an antineutrino.  

\begin{figure}
    \centering
    \includegraphics[width=0.8\textwidth]{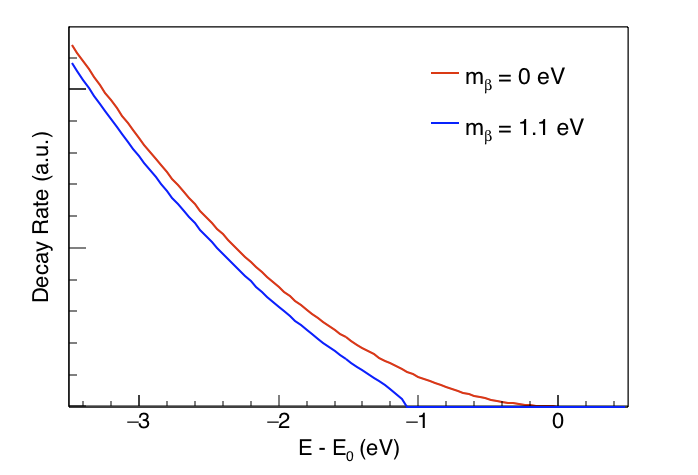}
    \caption{The atomic tritium endpoint spectrum under different neutrino mass scenarios both having the same extrapolated endpoint energy.}
    \label{fig:tritium_endpoint}
\end{figure}

All of the neutrino mass information is contained near the endpoint energy of the decay.  Therefore, to a good approximation, the differential electron energy ($E$) spectrum near the endpoint can be modeled as~\cite{Doe:2013fk}:

\begin{equation}\label{eqn:tritiumBetaDecay}
\frac{dN}{d\epsilon} = 3rt \epsilon \sqrt{\epsilon^2 - m_{\beta}^2} \approx 3rt \epsilon^2 \left( 1 - \frac{m_{\beta}^2}{2 \epsilon^2} \right),
\end{equation}

\noindent where $\epsilon \equiv E_0 - E$ ($E_0$ is the endpoint), $r$ is the rate into the last 1\,eV of the spectrum with $m_\beta =0$, and $t$ is the observation time. An example of the endpoint spectrum distortion due to a finite neutrino mass is shown in Figure~\ref{fig:tritium_endpoint}.  The current best limit on the neutrino mass using this method is from the KATRIN experiment, with $m_\beta \le $\SI{0.8}{\electronvolt} mass~\cite{bib:KATRINNature} at 90\% C.L..

The simple form of Equation~\ref{eqn:tritiumBetaDecay} belies the extreme difficulty of a tritium endpoint measurement.  Besides the obvious requirement for very good energy resolution ($\Delta E \sim m_\beta$ at $E_0 = 18.6$\,keV), the statistical sensitivity has terrible scaling relationships that tend to result in huge spectrometers, and the natural form of tritium gas, molecular T$_2$, has an irreducible systematic associated with final states~\cite{Bodine:2015aa}.  The first unfavorable scaling relation follows from the extreme rarity of events near the endpoint; $2 \times 10^{-13}$ of all events occur in the last 1\,eV.  Therefore, large amounts of tritium are required to gather sufficient events near the endpoint, while virtually all of the spectrum below the endpoint contributes nothing to the neutrino mass extraction.  The scaling challenge is exacerbated by the fact that the observable is $m_\beta^2$; for an order of magnitude improvement in $m_\beta$ sensitivity, we actually need 4--5  orders of magnitude increase in statistical sensitivity alone, not to mention commensurate improvements in systematics.  The large size of tritium endpoint experiments follows in part from the need to accommodate sufficient tritium source exposure for statistical sensitivity at the endpoint, and to manage the uninteresting low-energy events.  The final-state systematic reflects uncertainty in the width of a narrow band of rotational and vibrational states of the \ce{^3HeT^+} daughter populated in the decay of molecular tritium.  The final-states spectrum introduces an irreducible systematic uncertainty in any experiment with a molecular tritium source.  Even with 1\%-precision knowledge of the final-state spectrum of the \ce{T2} molecule, a molecular source limits sensitivity to about \SI{0.1}{\electronvolt}.

The molecular final states and their uncertainties are just one example of difficult systematic challenges in the tritium endpoint method. In addition, Robertson and Knapp~\cite{Robertson:1988aa} identify three other effects: the resolution function of the electron spectrometer, electron energy loss in the source, and backgrounds. Each one of these affects the shape of the tritium endpoint in the same way as neutrino mass. Furthermore, an uncertainty in the background rate or the width of the instrumental response produces an additional systematic error in the neutrino mass without affecting the quality of the fit. These quantities must be both small and precisely known.

Electron spectroscopy in current state-of-the-art tritium-endpoint experiments is performed by magnetic adiabatic collimation with an electrostatic (MAC-E) filter~\cite{Lobashev:1985mu}. A MAC-E filter measures the integral of the spectrum above the electrical potential in its analyzing plane; the differential spectrum must be constructed by scanning the potential.  KATRIN's spectrometer is of MAC-E type. % (See Figure~\ref{fig:KATRIN}).
%\begin{figure}
%    \centering
%    \includegraphics[width=\textwidth]{figures/KATRIN.png}
%    \caption{The Karlsruhe Tritium Neutrino Experiment (KATRIN).}
%    \label{fig:KATRIN}
%\end{figure}
KATRIN's source has the maximum tolerable column density consistent with the requirement to transport electrons to the main spectrometer without significant probability of scattering.  The source therefore cannot be made more dense or longer along the direction of the magnetic field.  The intensity can only increase by enlarging the source radially, with proportional radial expansion of the main spectrometer.  KATRIN's main spectrometer is already 10\,m in diameter and maintained at \SI{e-11}{\milli\bar} pressure~\cite{Aker_2021}.  An order-of-magnitude increase in sensitivity would require a spectrometer at least \SI{300}{\meter} in diameter at the same pressure.  This extreme technical feat would be futile without a similar order-of-magnitude reduction in systematic uncertainties, motivating the search for a new technique to push beyond KATRIN's projected \SI{0.2}{\electronvolt} sensitivity limit.\\

\subsection{Cyclotron Radiation Emission Spectroscopy}\label{sec:CRES}
A new method for electron spectroscopy that could avoid the technical limits of MAC-E spectrometry, Cyclotron Radiation Emission Spectroscopy (CRES), has been proposed~\cite{Monreal:2009za} and demonstrated~\cite{Project8:2014ivu,Project8:2017nal}.  In CRES, a gas source decays in a magnetic field $B$.  Emitted electrons trace cyclotron trajectories along $B$-field lines.  The centripetal acceleration results in the emission of coherent radiation at the cyclotron frequency~$f$:
\begin{equation}\label{eqn:f_cyclotron}
2 \pi f = \frac{2 \pi f_0}{\gamma} = \frac{eB}{m_e + E/c^2}
\end{equation}
where $\gamma$ is the Lorentz factor, $e$ is the fundamental charge, $m_e$ is the mass of the electron, $c$ is the speed of light, and $E$ is the energy of the electron.
The zero-energy (non-relativistic limit) electron cyclotron frequency $f_0$ is a fundamental constant \cite{Zyla:2020zbs},
\begin{eqnarray}
f_0&=& 27.992489872(8) {\rm \ GHz\ T}^{-1}.
\end{eqnarray}

The free-space emitted cyclotron power follows the Larmor formula,
\begin{equation}
P = \frac{2 \pi e^2 f_0^2}{3 \epsilon_0 c} \frac{\beta^2 \sin^2\theta}{1-\beta^2}.
\end{equation}
The pitch angle $\theta$ is the angle between the momentum vector and field direction.
The maximum power radiated by an 18-keV electron in a 1-T field is about 1 fW.

\begin{figure}[htb]
\begin{center}
\includegraphics[width=0.85\textwidth]{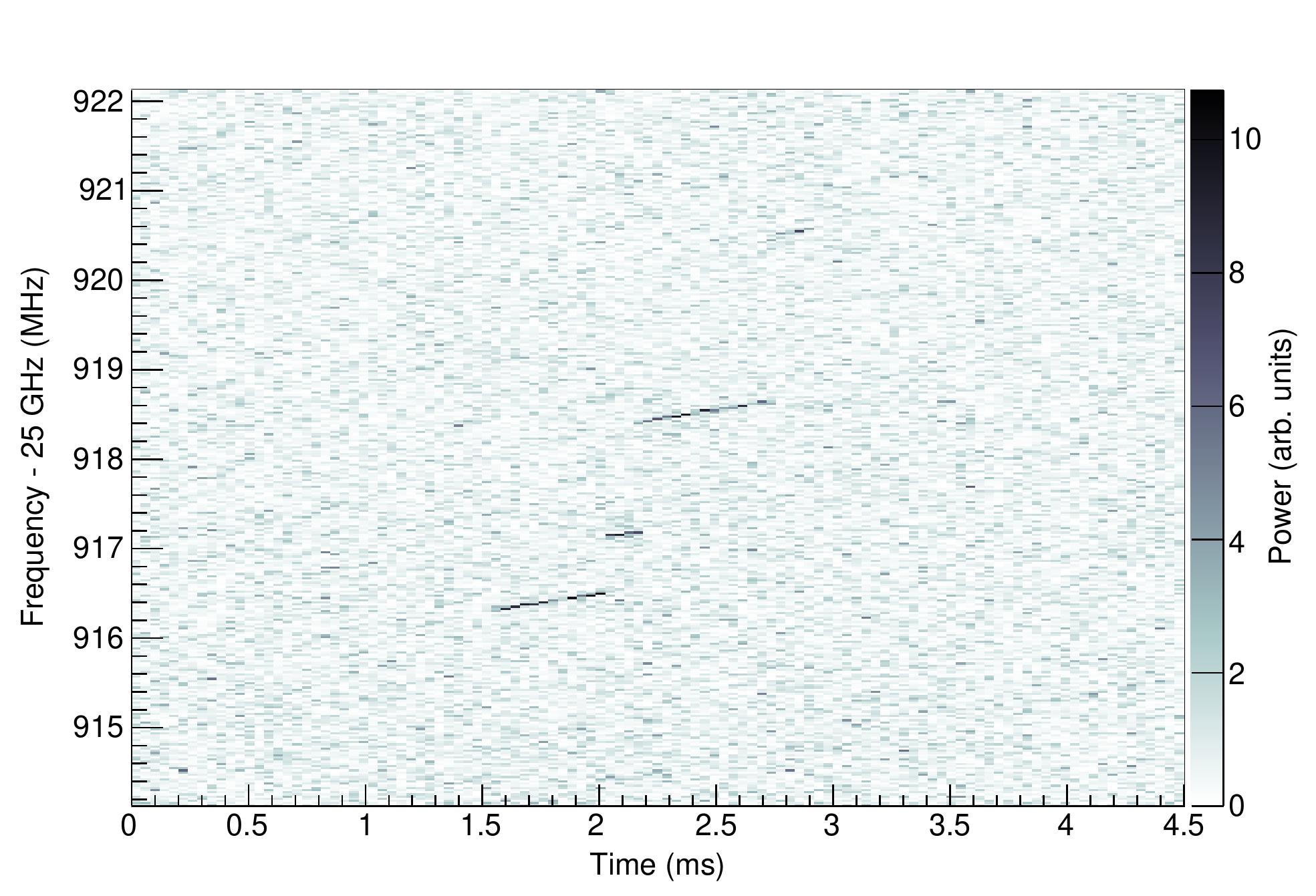}
\caption{\label{fig:CRES_event} A single tritium beta-decay electron recorded with the CRES technique. An electron is created by \isotope{H_2}{3} decay near the lower left corner and forms a track that slopes upward due to radiation loss.  The discontinuities from track to track are caused by the electron scattering inelastically from the residual tritium gas.  The most probable jump size corresponds to about 14 eV. Eventually the electron scatters out of the trap and is lost.}
\end{center}
\end{figure}

The cyclotron frequency depends on the kinetic energy, so a measurement of an electron's cyclotron frequency is tantamount to a measurement of its energy. A typical CRES event is shown in Figure~\ref{fig:CRES_event}.  The event begins abruptly, chirps toward higher frequency (lower energy) as it radiates, and makes a series of frequency (energy) hops as it suffers discrete scatters on residual gas in the high-vacuum environment. The start frequency of the first high-power segment, or ``track'', encodes the kinetic energy of the electron at the instant of its emission. The total path length of the electron in Figure~\ref{fig:CRES_event} is about \SI{60}{\kilo\meter}.  A magnetic trap is therefore required to keep electrons in the sensitive volume of a CRES detector. Trapping introduces a dependence of the frequency on the pitch angle $\theta$ between the electron momentum and the magnetic field. The value of $B$ in Equation~\ref{eqn:f_cyclotron} is the average value of the field experienced by an electron during an observation interval.  Electrons with smaller pitch angles (more nearly parallel to $\vec{B}$) explore larger variations from the minimum field to reach their turning points, shifting their observed frequencies upwards~\cite{AshtariEsfahani:2019yva}.

The merits of CRES are the extreme precision and accuracy possible with a nondestructive frequency-domain technique combined with very low backgrounds, and the fact that a tritium gas source is transparent to the microwave-frequency signal.  The latter benefit enables the source region to be surrounded by instrumentation. Uninteresting lower-energy (higher-frequency) beta decay electrons can be removed with simple low-pass filtering of the recorded signal. This is expected to result in a much more favorable scaling relation than for MAC-E spectrometers.  Furthermore, CRES measures the entire differential endpoint spectrum simultaneously with significant advantage over integrating spectrometers for statistical sensitivity.  Simultaneous measurement of the differential spectrum also avoids systematic uncertainties related to the stability of the source and spectrometer that are inherent to a scanned integral spectrum measurement.

Background events are expected to be extremely rare for the CRES technique.  We distinguish two classes of background: physics backgrounds and false triggers.  Physics backgrounds occur when an electron from a source other than tritium beta decay becomes trapped.  To be mistaken for a real signal event, the electron must have similar kinetic energy and a momentum consistent with being trapped (momentum nearly perpendicular to the field).  Beta-decaying gas contaminants are negligible in the vacuum environment.  Alpha particles from gas contaminants would appear at far lower powers and frequencies even if they weren't negligibly rare.  Electrons emitted due to cosmic ray interactions in the vessel walls simply return immediately to the walls because of the high magnetic field. Only electrons created by cosmic ray interaction on the gas can become trapped. This physics background in a large CRES experiment has been reliably estimated based on similar calculations of backgrounds in KATRIN. Even for extremely large sources up to order \SI{100}{\meter^3} the rate is only of order 1 event/y/eV.

False trigger backgrounds occur when noise fluctuations in neighboring pixels of a spectrogram like Figure~\ref{fig:CRES_event} conspire by chance to create a feature that appears to analysis algorithms like a real track.  These can be made rare, especially with good signal-to-noise ratio (SNR).  By choosing a threshold for total power integrated along a spectrogram track, the false-event rates can be made arbitrarily small, at the expense of detection efficiency for short tracks from electrons that quickly scatter.  

The combination of these favorable factors (low background, high precision) make CRES an attractive option for pushing the neutrino mass sensitivity through beta decay.  Section~\ref{sec:Project8} discusses recent achievements in using CRES for electron spectroscopy, while Sections~\ref{sec:PhaseIII} and~\ref{sec:atomic} describe planned R\&D efforts of the collaboration in reaching its ultimate sensitivity goal of $m_\beta \ge 40$~meV/c$^2$. A discussion of the projected sensitivity of the CRES technique is discussed in Section~\ref{sec:Sensitivity}.
\section{The Project 8 Experiment}\label{sec:Project8}

{\em The Project 8 neutrino mass experiment makes use of the CRES technique in order to make a direct neutrino mass measurement with a final projected sensitivity of 40 meV/c$^2$ at 90\% confidence level.  The collaboration at the time of this writing consists of approximately 50 scientists across the United States and Germany.  The collaboration has approached these measurements through technology demonstrations, internally referred to as ``Phases.''  Phase~I provided a proof-of-principle of the CRES technique by using a gaseous krypton source, $^{83m}$Kr, while Phase~II provided a first demonstration of the CRES technique using molecular tritium. This section summarizes the results from those two measurement programs.}

\subsection{Phase I: Kr Demonstrator Results}
\label{sec:phaseI}

\begin{figure}[htb]
  \begin{center}
  \begin{tabular}{c c}
  \includegraphics[width=0.70\textwidth]{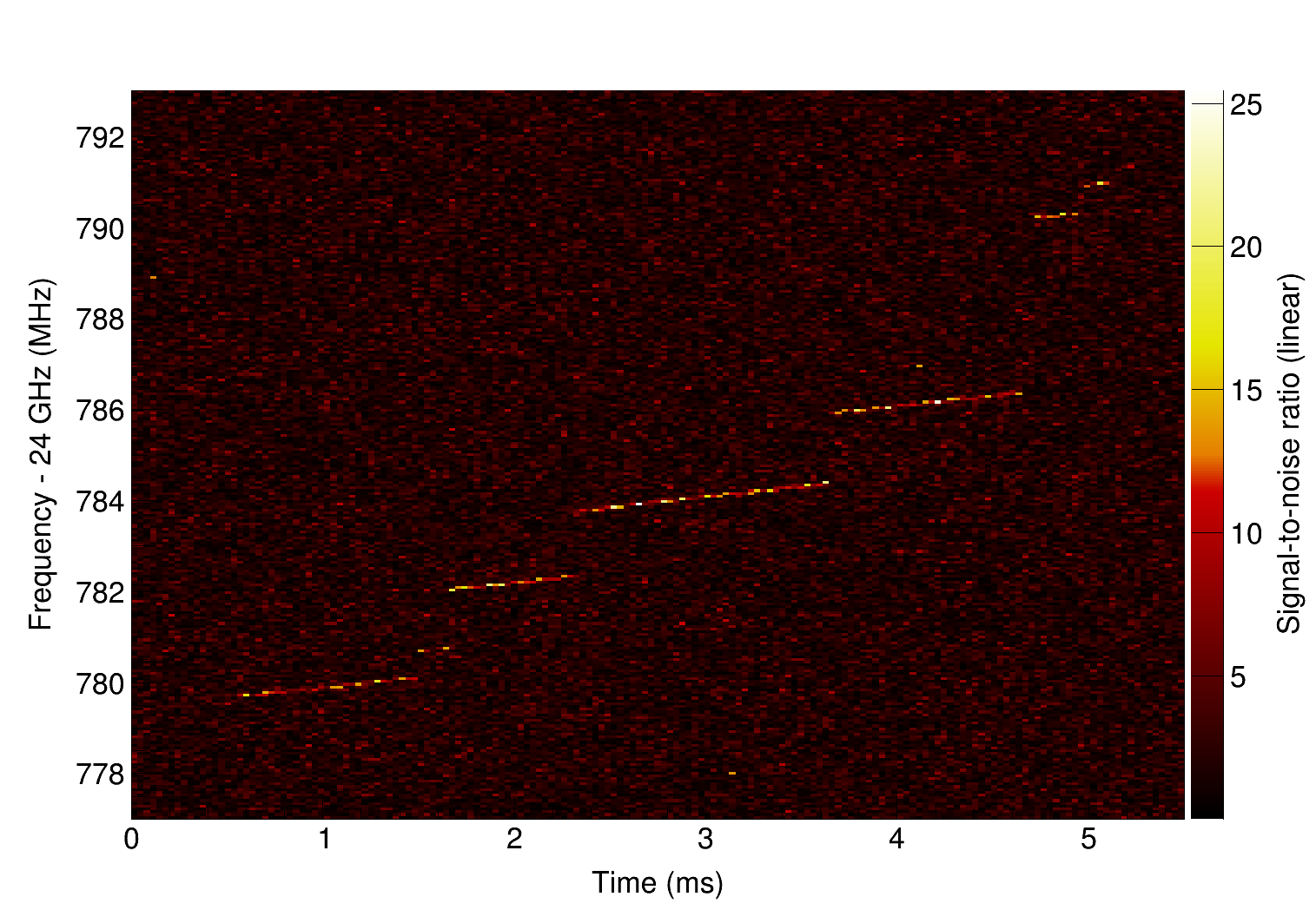} &
  \includegraphics[width=0.12\textwidth]{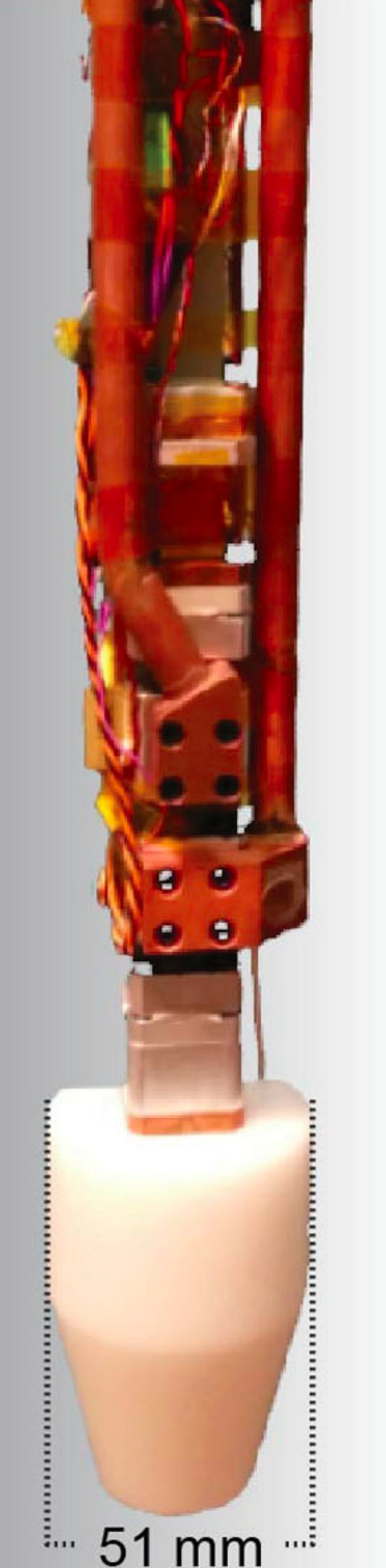} \\
  \end{tabular}
  \caption{Left: First event observed by the CRES method (Project 8 Collaboration \cite{Project8:2014ivu}). The spectrogram shows RF power in 25-kHz frequency bins and 40-$\mu$s time bins.  An electron is created by $^{83m}$Kr decay near the lower left corner and forms a track that slopes upward due to radiation loss.  The discontinuities from track to track are caused by the electron scattering inelastically from the background gas, which is mainly hydrogen.  The most probable jump size corresponds to about 14\,eV. Eventually the electron scatters out of the trap and is lost. Right: Waveguide insert used for Phase I of the measurement.}
  \label{fig:eventzero}
  \end{center}
\end{figure}

The first experimental demonstration of CRES was made in 2014 using the isotope $^{83m}$Kr, the decay of which produces internal conversion electrons (CE).  The radioactive isotope is a gamma-emitting isomer of \isotope{Kr}{83} with a half-life of 1.8\,h, in which internal conversion produces mono-energetic electron lines with kinetic energies of 17824.2(5)\,eV, 30226.8(9)\,eV, 30419.5(5)\,eV, 30472.2(5)\,eV, 31929.3(5)\,eV, and 31936.9(5)\,eV, with line widths less than 4\,eV~\cite{Venos:2018aa}. The experimental cell consisted of a section of WR-42 rectangular waveguide with a cross section of $10.7 \times 5.0$\,mm.  Because the electrons  travel a great distance in the several microseconds needed to make an accurate measurement of the frequency, a magnetic trap formed by a coil around the waveguide was used to trap electrons having pitch angles $\theta$ near $\pi/2$.  Signals produced by electrons were transmitted by the waveguide to a low-noise cryogenic amplifier, superheterodyne receiver, and digitizer. The first event recorded is shown in the iconic plot reproduced in Fig.~\ref{fig:eventzero}.  

\begin{figure}[htb]
  \begin{center}
  \begin{tabular}{c c}
  \includegraphics[width=2.7in]{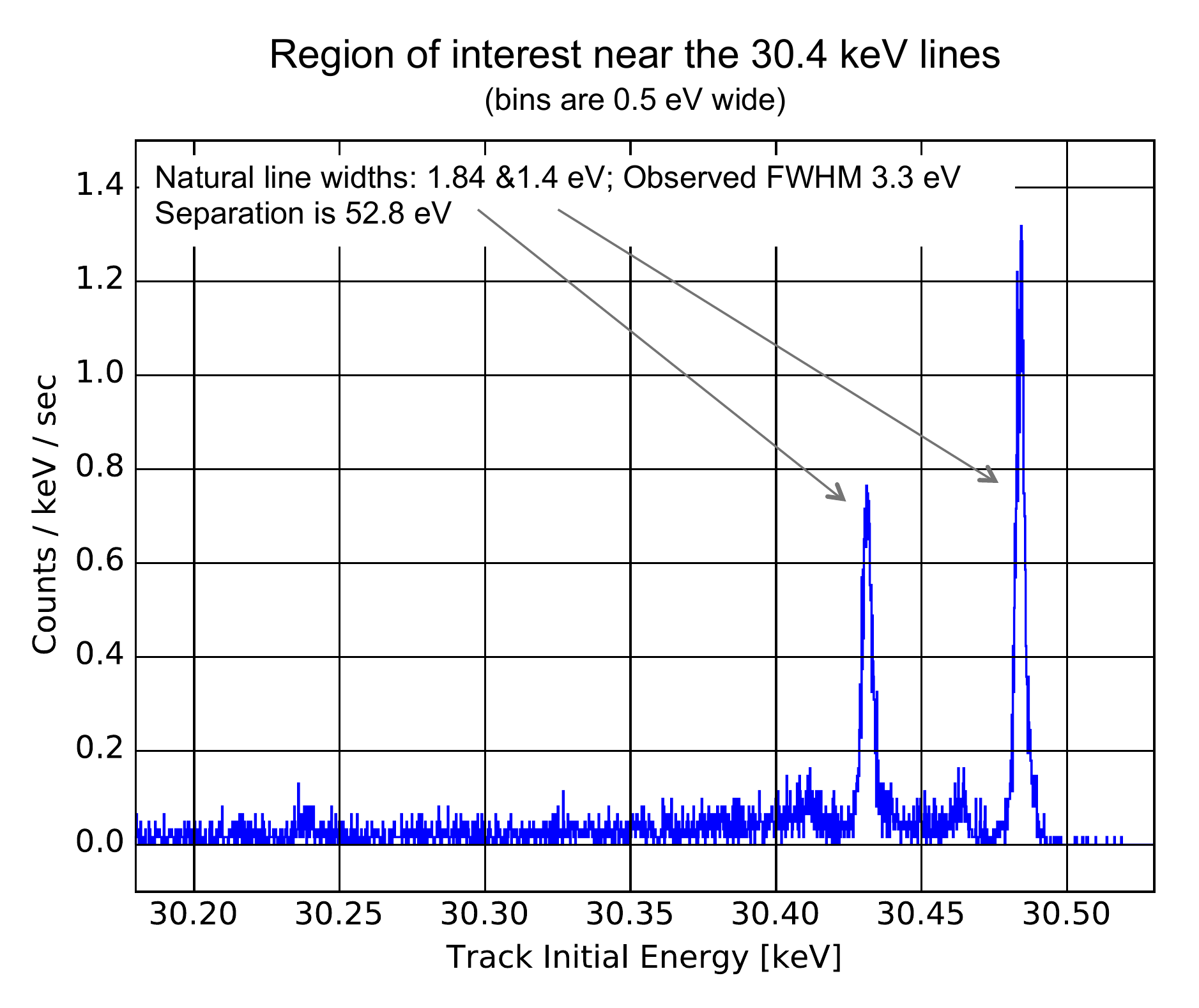} &
  \includegraphics[width=2.7in]{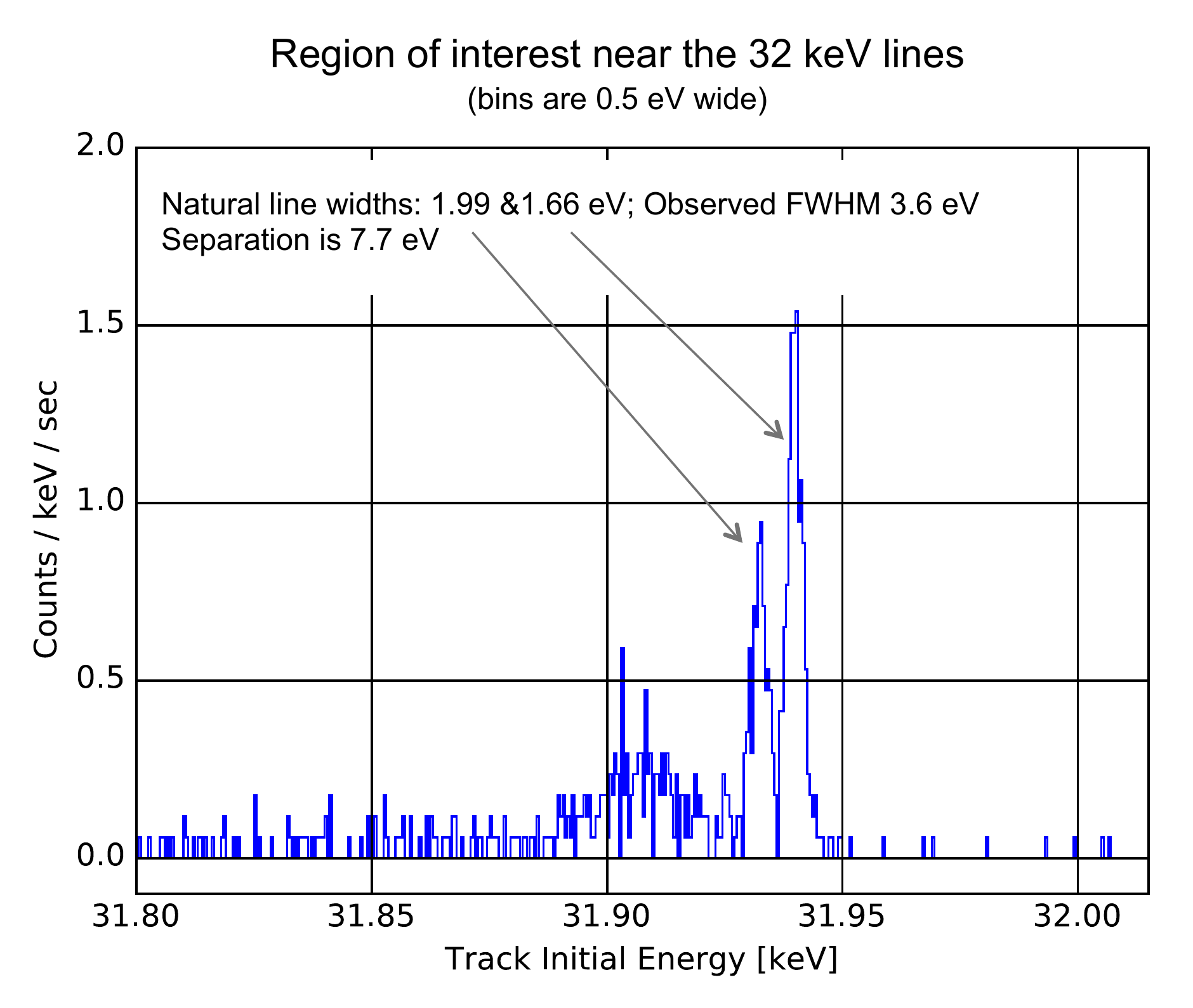} \\
  \end{tabular}
  \caption{Internal-conversion electron lines in the decay of $^{83m}$Kr measured by the Project 8 collaboration with the CRES method \cite{Project8:2017nal}. Left: The L2 and L3 lines from the 32\,keV isomeric decay transition. Right: The M2 and M3 lines from the same transition.  The events not in the sharp peaks arise mainly from shakeup and shakeoff processes in the decay \cite{Robertson:2020boa}, and partly from scattering in the residual gas.}
  \label{fig:krlines}
  \end{center}
  \end{figure}

For each such  decay event, the initial electron energy is derived from the frequency at the onset of power in the first track. Other features characteristic of the CRES signature, including the energy loss due to cyclotron radiation (slope in frequency) and the sudden changes in frequency due to inelastic collisions (``jumps'' in frequency) are readily visible. Examples of the good resolution obtainable with the CRES method are shown in Fig.~\ref{fig:krlines}, taken from Ref.~\cite{Project8:2017nal}. The instrumental resolution is about 3 eV FWHM for these data.
  
The results from the Phase I demonstration were published in~\cite{Project8:2014ivu}.  The apparatus was upgraded later to accommodate gaseous molecular tritium, allowing a first demonstration of the CRES technique on beta decay (Phase~II).

\subsection{Phase II: Molecular Tritium Results}\label{sec:phaseII}

Phase~II demonstrated the capabilities of CRES for measuring continuous spectra by making the first-ever measurement of the molecular-tritium spectrum endpoint by this method. Molecular tritium was introduced into the system using a sintered porous non-evaporable getter and released using a feedback-loop-controlled heater.
The tritium operating pressure was stabilized at $10^{-6}$\,mbar, which optimally balances the event rate with the rate of unwanted collisions of electrons with the gas.  The $^{83m}$Kr gas could still be introduced into the system as a calibration electron source, akin to what was done for the Phase~I program.

\begin{figure}[htb]
    \centering
    \includegraphics[height=0.28\textheight]{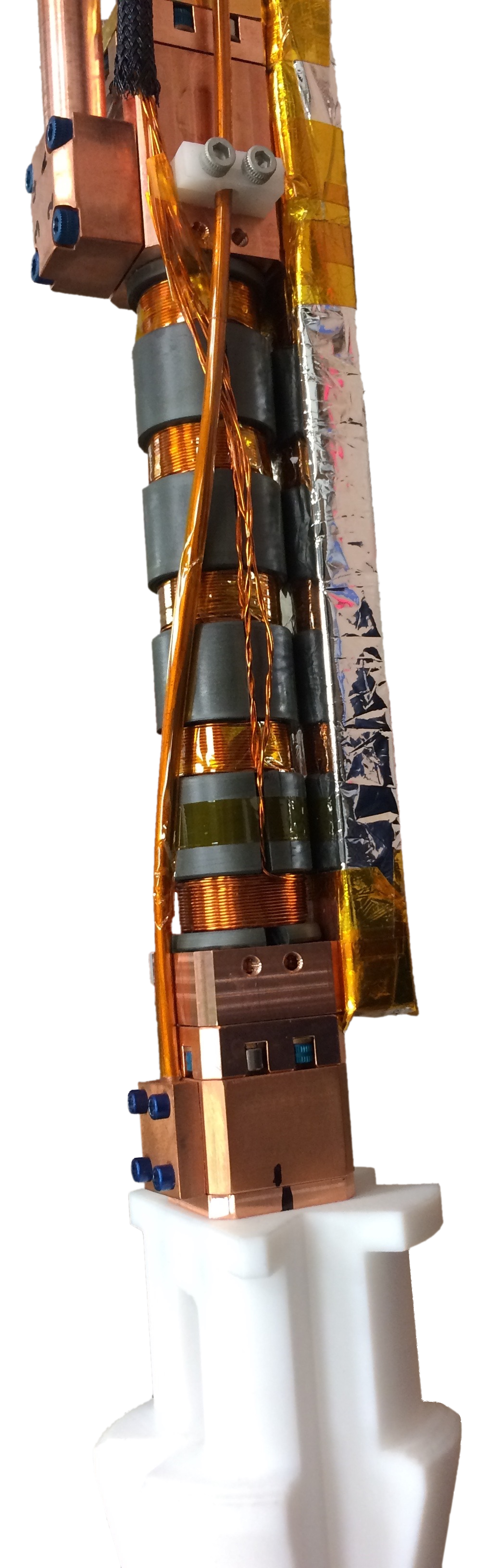}
    \includegraphics[width=0.40\textwidth]{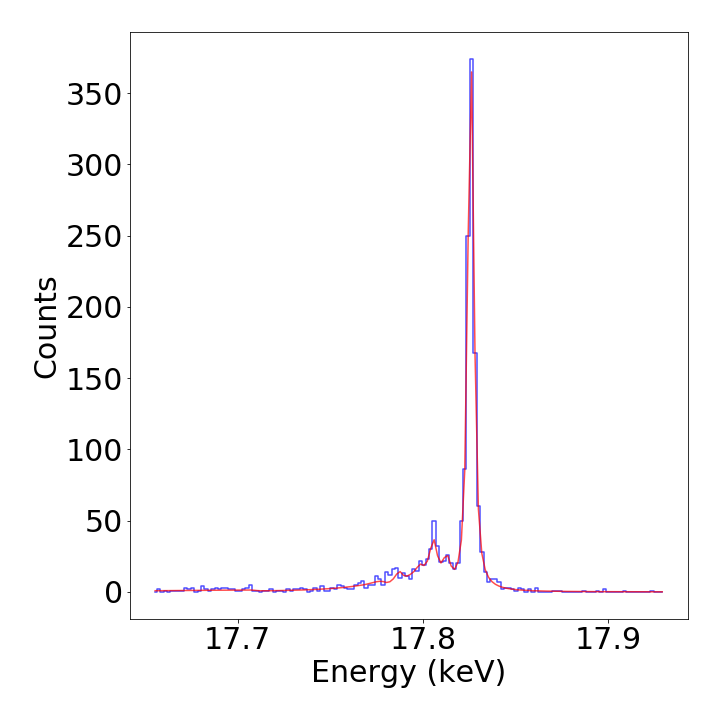}
    \includegraphics[width=0.45\textwidth]{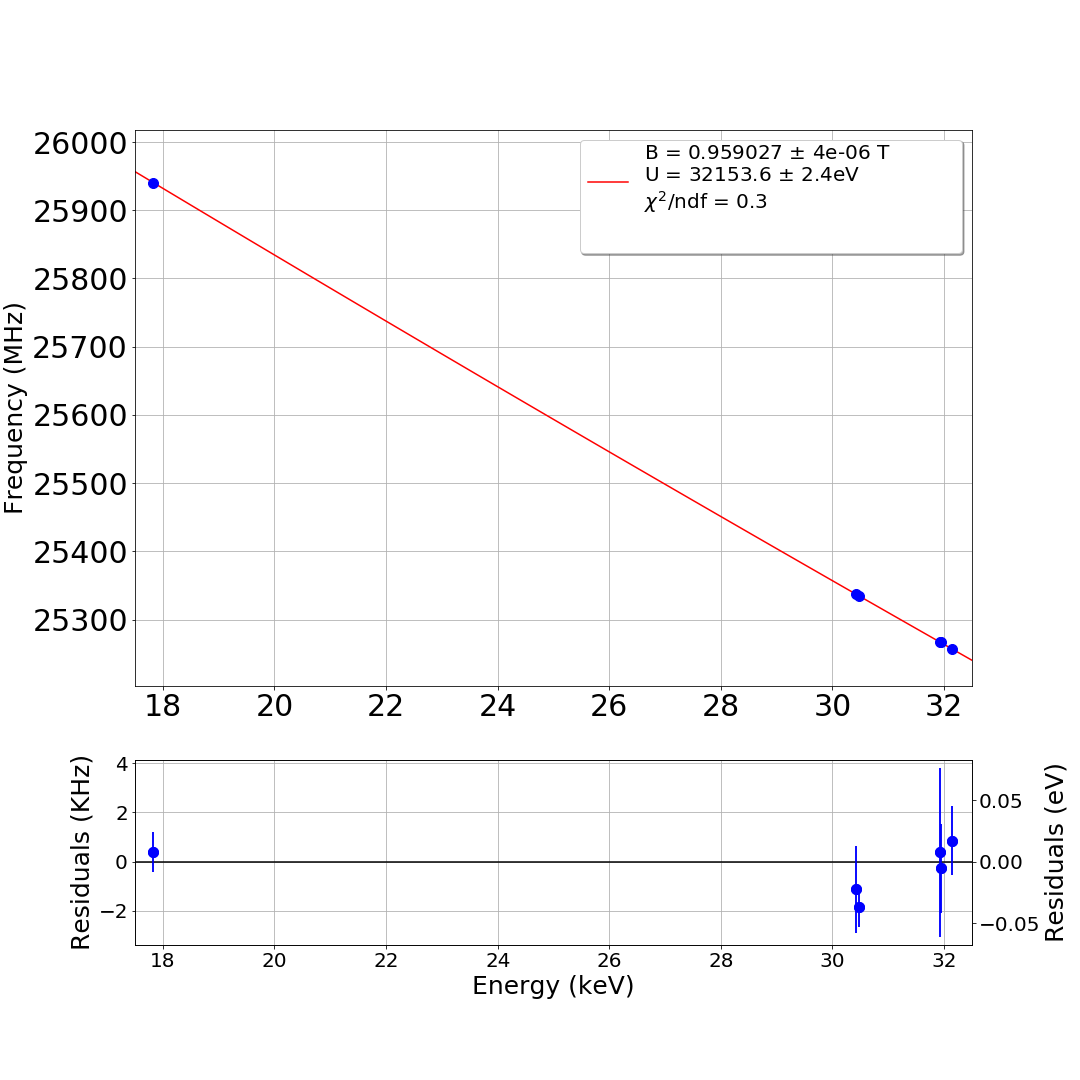}
    \caption{Left: The Phase~II CRES cylindrical waveguide insert. Middle: Measured \ce{^{83{\rm m}}Kr} K-Shell CE fitted with model including shake-up, shake-off, and electron collisions with hydrogen molecules and krypton atoms. Right: Measured frequency versus energy of the K, L and M-shell \ce{^{83{\rm m}}Kr} CEs.}
    \label{fig:PhaseII_insert}
\end{figure}

The Phase~II insert is shown with measured calibration \isotope{Kr}{83m} CE results in Figure~\ref{fig:PhaseII_insert}. The middle plot is the measured \ce{^{83{\rm m}}Kr} K-shell CE. The red curve is a fitted model accounting for both the  natural line width and instrumental artifacts. The instrumental resolution is \SI{1.73}{\electronvolt} at FWHM.  The right plot shows the precision of the conversion between measured cyclotron frequencies, and the known energies of the K, L, and M-shell conversion electrons.  The tritium endpoint method is based on measuring the shape of the tritium decay spectrum; any non-linearity that distorts that shape will lead to systematic uncertainty.  The region of interest for the tritium endpoint is just a few 10s of eV at most, whereas Phase~II data demonstrates linearity over a much larger range of \SI{14}{\kilo\electronvolt}. 
\begin{figure}[hbt]
  \centering
  \includegraphics[width=0.7\columnwidth]{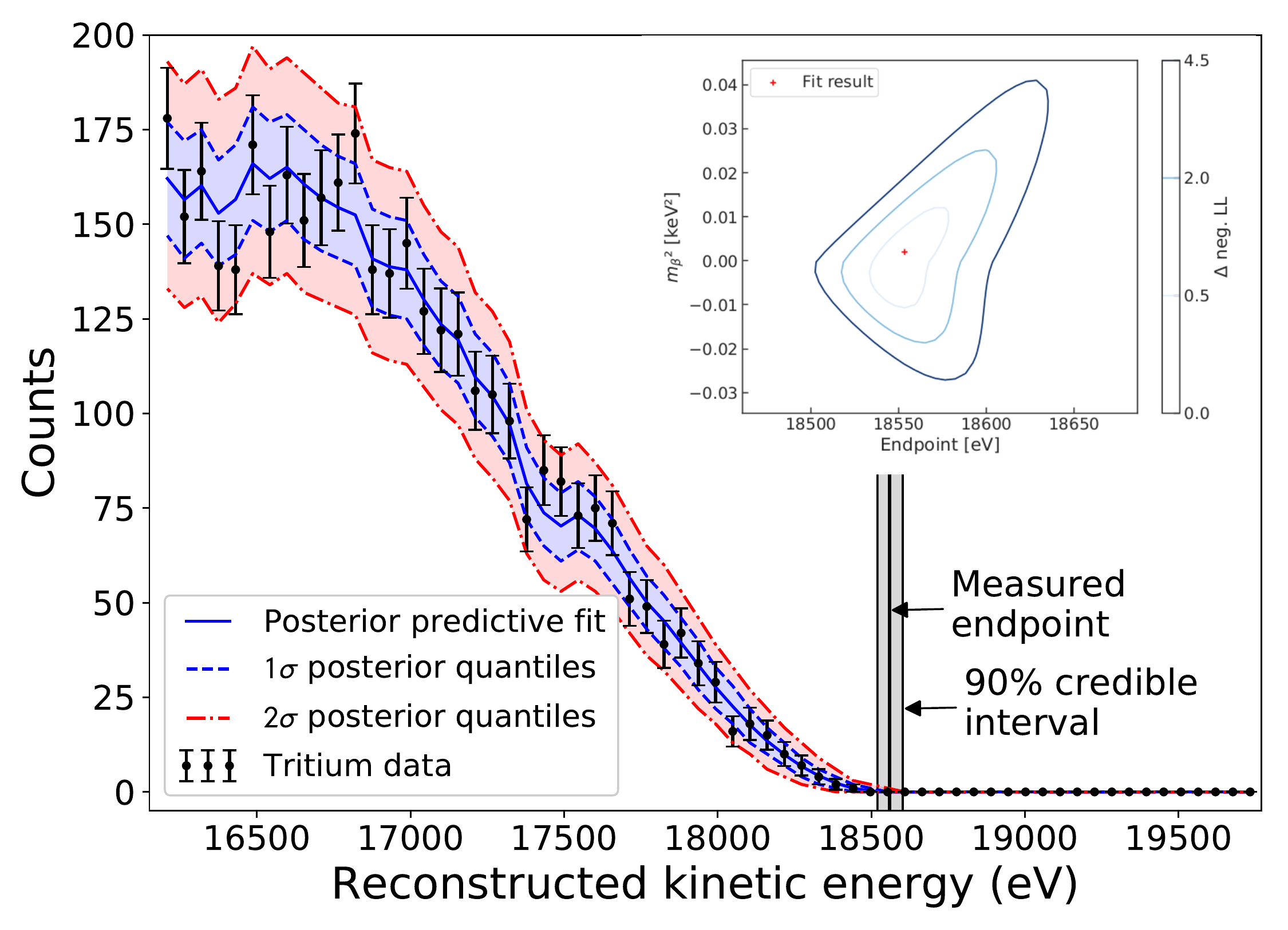}
  \caption{
    Measured tritium endpoint spectrum with posterior predictive from Bayesian analysis overlaid. (Inset) Neutrino mass and endpoint contours from the maximum likelihood fit to the spectrum.}
  \label{fig:phaseII_spectrum}
\end{figure}

Tritium data were taken over 82 days in the deep-trap configuration, detecting 3742 events. The analysis window spanned 25.81-25.99\,GHz, or 16.2-19.8\,keV in electron energy.  Figure~\ref{fig:phaseII_spectrum} shows the measured tritium endpoint spectrum and a model of the expected shape accounting for detection efficiency.  The measured endpoint value, $18559(25)$\,\si{\electronvolt}, is in good agreement with expectations~\cite{Bodine:2015aa}.  By measuring an interval that includes the region above the endpoint, and observing no events, a limit was set on the background rate of less than \SI{3e-10}{\per\electronvolt\per\second} (90\% confidence interval).  

Phase~II represents the first upper limit on the neutrino mass using CRES, which is not competitive but serves as a proof-of-principle. Along with the stringent background limit and the high resolution demonstrated in the shallow trap, these data support the claim that the CRES methodology has the potential to extend its sensitivity to the neutrino mass scale.

\subsection{Phenomenology and Simulations}
\label{sec:phaseIII}

\begin{figure}[htb] 
    \centering
    \includegraphics[width=0.40\textwidth,angle=90]{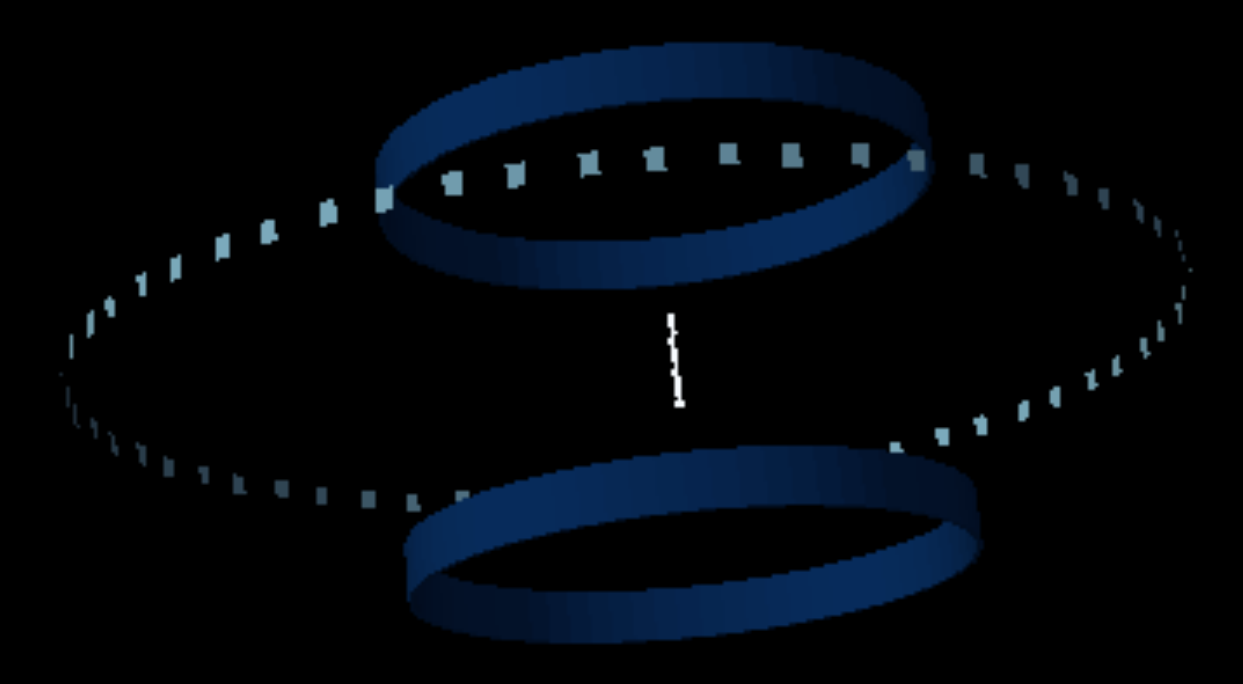} \\
     \begin{tabular}{c c}
         \includegraphics[width=0.38\textwidth]{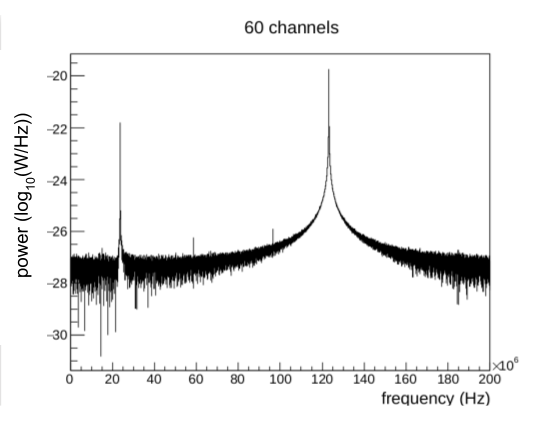} &
        \includegraphics[width=0.42\textwidth]{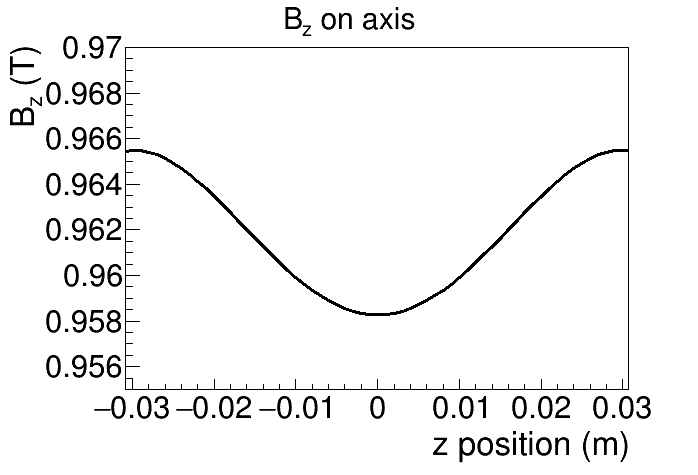} 
     \end{tabular}
     \caption{Top: A Locust/Kassiopeia simulation of an electron trapped at the center of the antenna ring of the Free Space CRES Demonstrator (see Section~\ref{sec:FSCD} for more details). The blue rings are electron trap coils, the antenna elements are represented as gray patches. The electron, emitted at $88^\circ$ with respect to the magnetic field, follows the path in the central white region.  Bottom: The received power spectrum (left) and the approximately harmonic magnetic field along the axis (right).}
     \label{fig:sim_prior_progress}
\end{figure}

\begin{figure}[hbt]
  \centering
  \includegraphics[width=0.65\columnwidth]{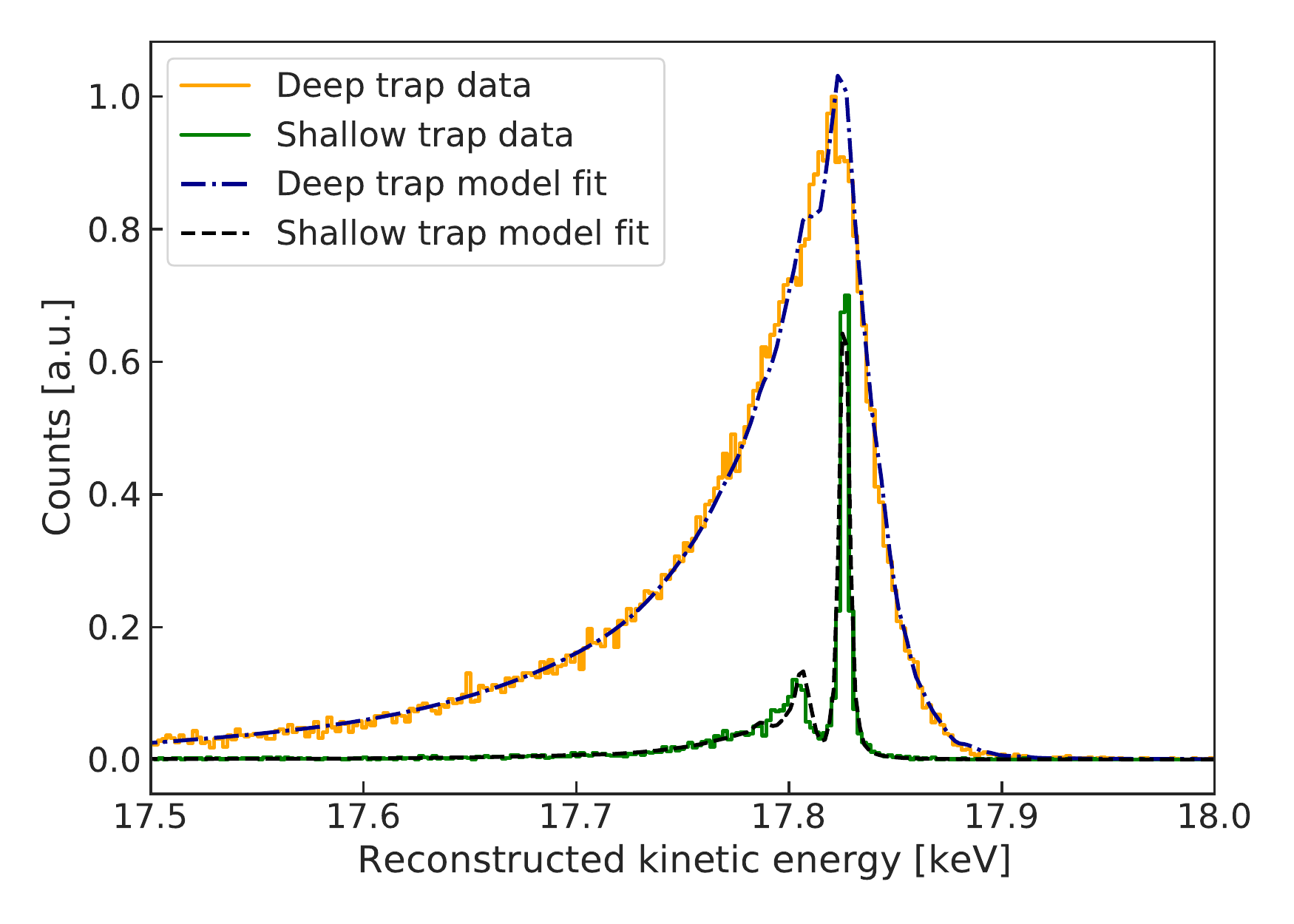}
  \caption{
    The 17.8 keV $^{83m}$Kr conversion electron K-line, as measured with CRES in the shallow (high-resolution) and the deep (high-statistics) electron trapping configurations. Deep and shallow trap fits incorporate simulated detector resolution response due to inelastic scattering, magnetic field broadening and signal-to-noise threshold effects.
  }
  \label{fig:resolution}
\end{figure}

Phases~I and~II of our measurement program have provided not only a proof-of-principle of the CRES technique for electron spectroscopy, but also a detailed understanding of the expected response of the method.  The data gathered from these measurement programs provide a detailed phenomenological understanding of CRES in a waveguide environment~\cite{AshtariEsfahani:2019yva}.  CRES phenomenology is characterized by an original simulation package called Locust~\cite{Esfahani:2019ab}.  To simulate electron trajectories, Locust uses KATRIN's Kassiopeia package~\cite{Furse:2017aa}, which can model static fields and the motions of particles in them.  Locust simulates the electron radiation fields, their propagation and collection by antennas, and the receiver response.  Figure~\ref{fig:sim_prior_progress} shows an example of a processed signal following a Locust/Kassiopeia simulation of one electron.  
These simulation and phenomenological tools perform extremely well when tested against data.  An example of this agreement comes in modeling the waveguide detector response, as shown in Fig.~\ref{fig:resolution}.
 We are using these simulation tools to support evaluations of mature conceptual design candidates that are likely to become the first large-volume demonstration of CRES technology.
\section{Future Demonstrators: Signal Detection and Source Development}
\label{sec:PhaseIII}

{\em Despite the remarkable progress made in the past few years in understanding the CRES technique and utilizing it for electron spectroscopy, considerable work remains to expand the technique for a large scale, inverted-order mass sensitivity experiment. In particular, the next stage has two major goals going forward:

\begin{enumerate}\bfseries
    \item Demonstrate that CRES is feasible and scalable in a large source volume while retaining high precision for electron energy reconstruction and maintaining low backgrounds.
    \item Demonstrate the ability to construct a high intensity atomic tritium source, so as to remove systematic uncertainties that pertain to molecular tritium operation.
\end{enumerate}
    
All these aspects are the focus of the next stage of R\&D for the experiment.  The effort culminates in a tritium endpoint experiment with sensitivity to limit $\boldsymbol{m_\beta \lesssim 1\,\si{\electronvolt}}$}.

\subsection{Large Volume Signal Detection}

One of the primary objectives of the next stage of the experiment is the ability to scale the CRES technique to be sensitive to beta decay electrons decaying in a high-density, large volume.  There exists a premium for both high reconstruction efficiency and energy resolution.  During the past two years, Project 8 has devised two possible options on how to achieve this:
\begin{itemize}
    \item {\em A Free Space CRES Demonstrator}, which uses a large array of high frequency ($\approx$ 26 GHz for a 1 Tesla magnetic field) antennas to reconstruct the energy and position of electrons in a given volume, and
    \item {\em A Cavity Resonator Design}, which uses a single cavity operated at low frequencies ($\sim$1 GHz) to reconstruct electrons.
\end{itemize}
The collaboration is presently evaluating these two technological approaches.  We present some of the details of each approach in the sections below. For both designs, the response to CRES signals will be evaluated at an intermediate volume and at $\approx 26$~GHz or a 1 T magnetic field first, given the availability of hospital-grade MRI magnets at this strength.

\subsection{Free Space CRES Demonstrator}
\label{sec:FSCD}

\begin{figure}[htb]
\begin{center}
\includegraphics[width=0.80\textwidth]{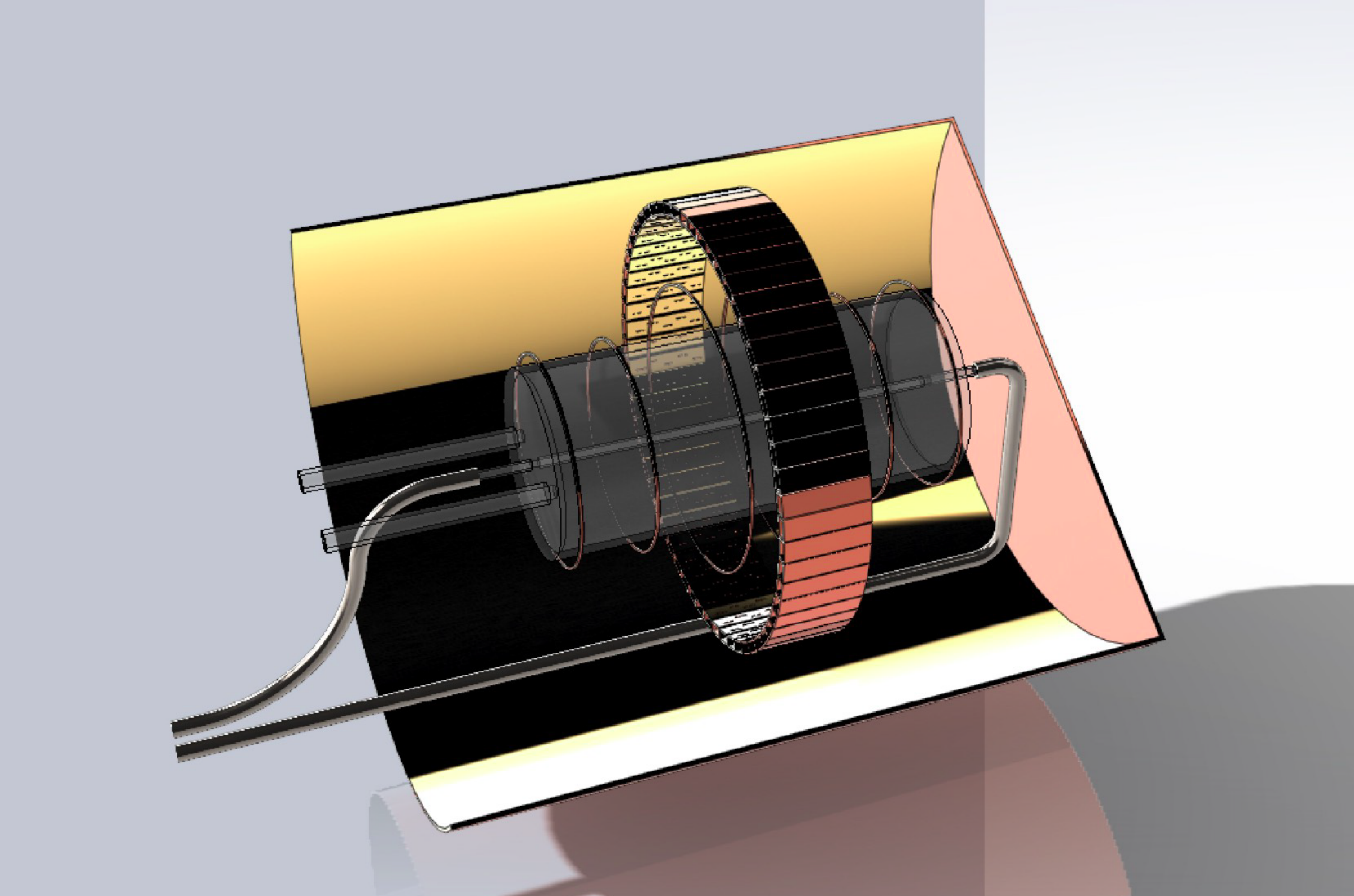}
\caption{\label{fig:phase_III_concept} A CRES detector concept using a free space antenna design optimized for high frequency (26 GHz). There is a uniform solenoidal magnetic field along the cylindrical axis. Current coils superimpose a magnetic trap that confines electrons within view of the antenna elements, arranged in a ring pattern at the center of the trap. The fiducial volume is about \SI{200}{\centi\meter^3}.}
\end{center}
\end{figure}

\begin{figure}[htb]
\begin{center}
\includegraphics[width=0.80\textwidth]{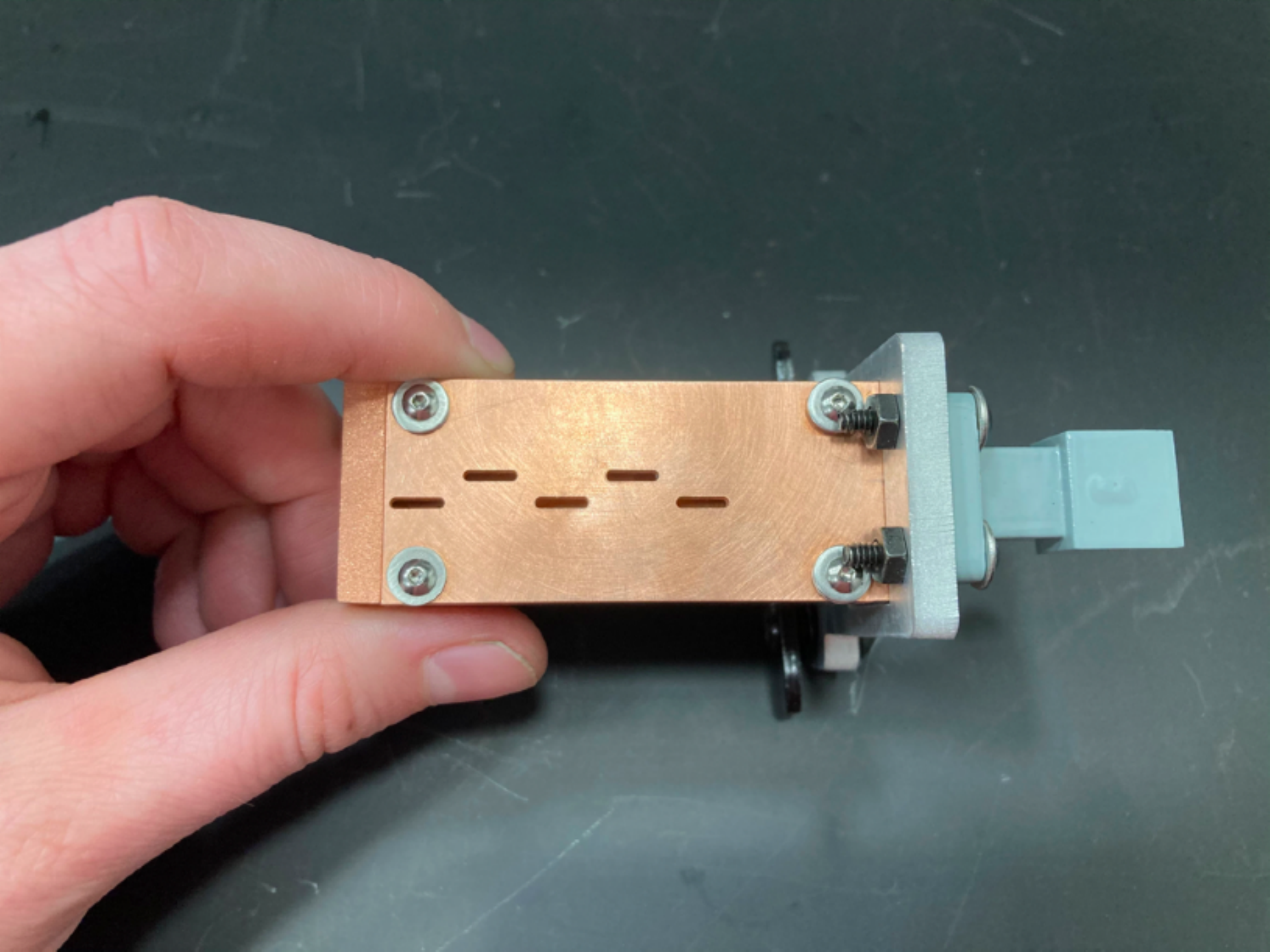}
\caption{\label{fig:singleantenna} A prototype slotted waveguide antenna for detection of 26 GHz microwaves.}
\end{center}
\end{figure}

The free space CRES Demonstrator consists of a cylindrical phased array of antenna elements aimed inwards at the electron source.  The antenna array is optimized to detect electron signals stemming from a 0.94 Tesla magnetic field (26 GHz frequency signal).  Internal studies have shown that using slotted waveguides for signal detection provide the highest signal-to-noise performance per channel (see Fig.~\ref{fig:singleantenna}).  To control noise, cost, and complexity, we consider configurations that reduce total channel count by passively combining elements and instrumenting them as groups. The conceptual rendering in Figure~\ref{fig:phase_III_concept} shows only individual elements with the grouping undefined. Leading design candidates combine elements passively along the cylindrical axis, but not around the azimuth.

\begin{figure}
    \centering
    \includegraphics[width=0.6\textwidth]{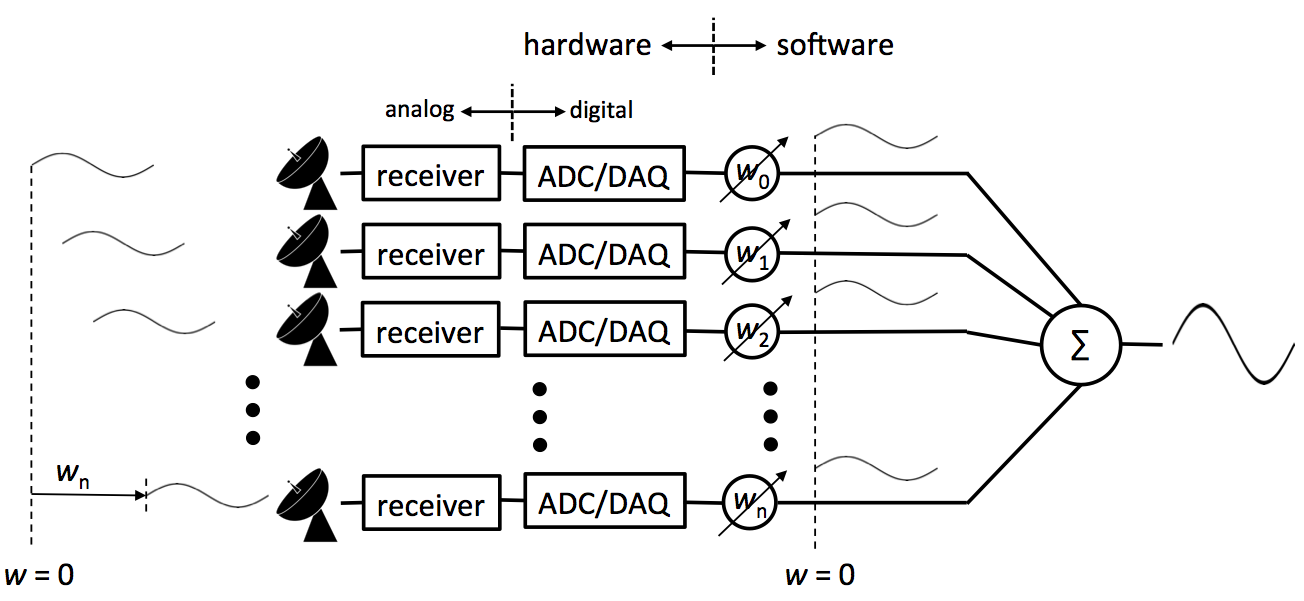}
    \includegraphics[width=0.3\textwidth]{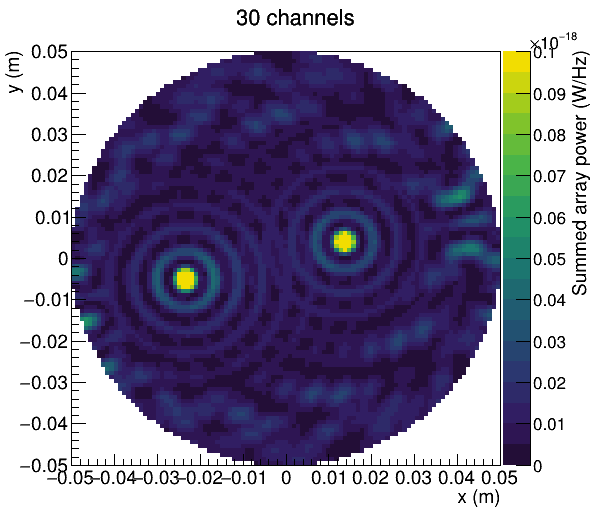}
    \caption{Left: The basic digital beam forming concept maps each focal point to a unique set of phase delays.  Right: Simulation of cyclotron radiation power received in a ring from two simultaneous electrons. The color represents received power spectral density versus the DBF focal position.}
    \label{fig:DBF}
  \end{figure}

Individually instrumented channels in the azimuthal direction permit digital beam forming (DBF) as shown on the left of Figure~\ref{fig:DBF}. In DBF, each possible focal point corresponds to a unique set of phase delays $w_i$.  Besides enforcing constructive interference of signal components received by different channels in a single ring, DBF provides localization in a narrow, roughly-cylindrical sensitive volume. That reduces systematic uncertainties following from inevitable radial variation in the magnetic field that shifts the cyclotron frequency according to Equation~\ref{eqn:f_cyclotron}, as long as a precise field map exists.  Figure~\ref{fig:DBF} (right) shows a simulated pileup event: the true locations of two electrons are apparent.

DBF increases the pileup-free event rate by allowing the system to record multiple electrons simultaneously as long as they are sufficiently separated in space.  Pileup resolution translates into demands on the data-acquisition and signal processing system discussed below.  Active DBF also allows the antenna array to ``follow'' electrons as their orbits slowly evolve due to small radial field gradients.

We have developed a recipe for calculating the positions, diameters and currents of trap coils to arrive at configurations ranging from harmonic traps that are parabolic around their minimum field, to almost arbitrarily flat fiducial volumes with sharp turning points referred to as a ``bathtub''. The ratio of the minimum to maximum field experienced by electrons determines what fraction is trapped:
\begin{equation}
\eta_{\rm trapped} = \sin\delta\theta_{\rm max}, \ {\rm where} \ \delta\theta_{\rm max} = \cos^{-1} \sqrt{\frac{B_{\rm min}}{B_{\rm max}}}
\end{equation}
is the range of trapped angles, measured relative to $\theta = \pi/2$.  For 10\% trapping efficiency, the ratio of fields is 99\%, corresponding to a range of trapped pitch angles $\delta\theta_{\rm max} = \pm 5.7^{\circ}$. 

Antenna-trap co-design is supported by Project 8's Locust simulation framework~\cite{Esfahani:2019ab} that describes the response of antennas to electromagnetic radiation. Locust generates realistic simulated data sets on which analysis approaches can be tested. Antenna responses are modeled with the commercial High Frequency Structure Simulation (HFSS) package~\cite{hfss}. KATRIN's Kassiopeia package~\cite{Furse:2017aa} handles calculations of static magnetic trapping fields and the precise tracking of electron motions within them. These are the tools to be validated by early FSCD data.

CRES antennas and the magnetic traps that confine electrons to their sensitive volume cannot be considered separately. Electrons move periodically through the spatially-varying gain profile of the antenna leading to amplitude modulation (AM) in the received signal.  Furthermore, radiation from relativistic tritium-endpoint electrons will appear alternately red- or blue-shifted at the antenna leading to frequency modulation (FM) of the received signals. Spatial magnetic field variations in the trap also contribute to FM. Both effects tend to displace power from the desired central cyclotron frequency $f$ into sidebands at $f \pm n \delta f$, where $\delta f$ is the frequency of modulation and $n$ is the harmonic order.

Detecting \SI{1}{\femto\watt} of intermittent power is challenging when it is all in the central cyclotron band; if that power is shared among even a few sidebands, signal-to-noise ratio (SNR) demands are even more severe. However, if sidebands are also detected, the pitch angle can be determined from $\delta f$ and used to correct the frequency shift discussed above, improving resolution.

Frequency modulation is particularly pernicious.  It can be quantified by a modulation index $h$, which in a CRES experiment turns out to be
\begin{equation}\label{eqn:FM_index}
h \sim \frac{\Delta f_{\rm Doppler}}{f_{\rm axial}},
\end{equation}
the ratio of the Doppler frequency shift to the axial frequency of motion in the trap.  For small indices, $h \lesssim 1$, most power is in the desired ``carrier'' band.  For larger values, $h\gtrsim 1$, most power is in sidebands with the unshifted central frequency disappearing altogether for some values. In an antenna-array CRES detector, $\Delta f_{\rm Doppler}$ can be reduced by moving the antennas to larger radii, and $f_{\rm axial}$ can be increased with shorter traps. FSCD conceptual design seeks the optimum relationship between antenna radii, beam pattern, and trap length.  The optimum maximizes detection efficiency.

\begin{figure}[hbt]
    \centering
    \includegraphics[width=0.53\textwidth]{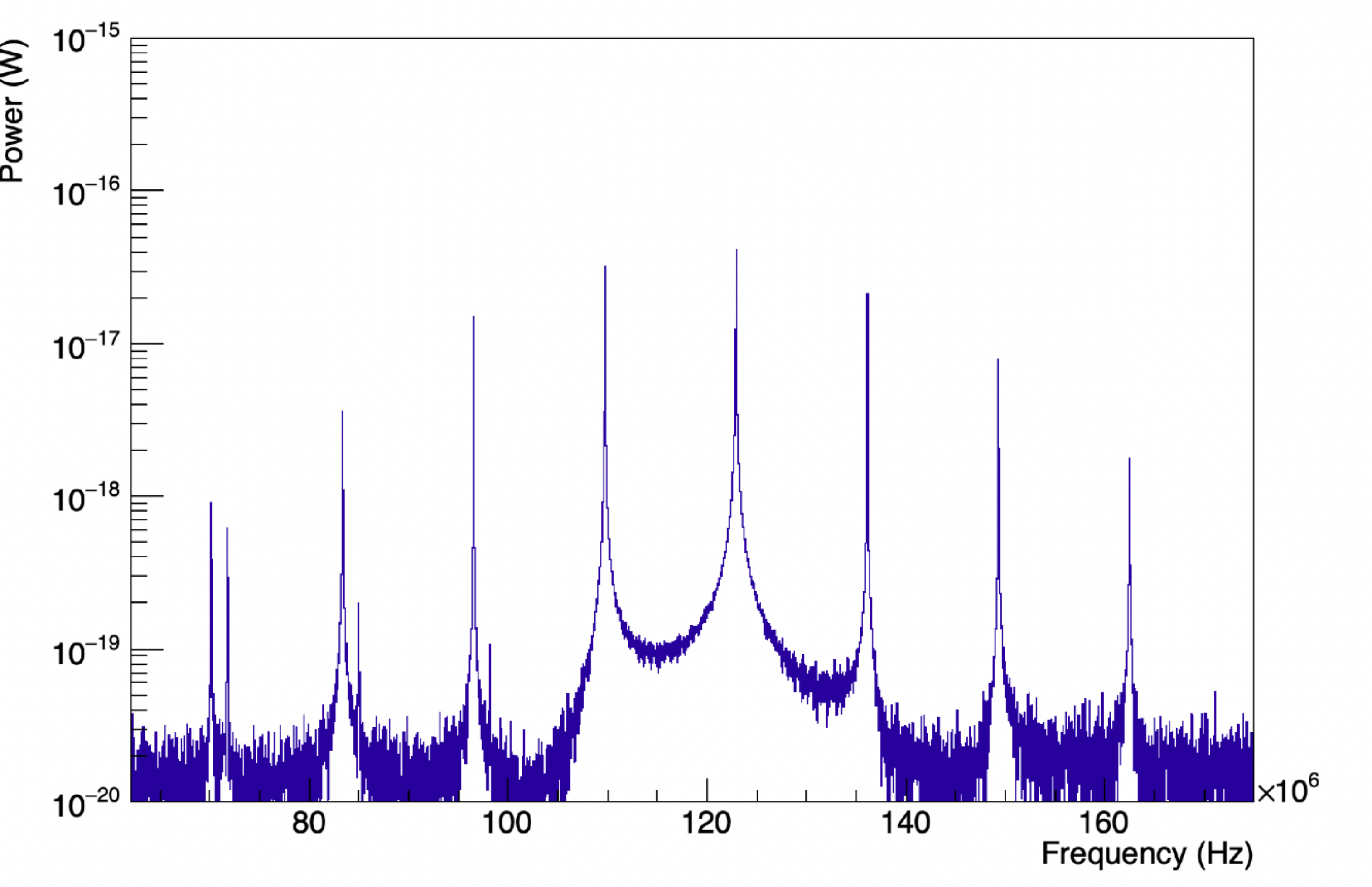}
    \hspace{2em}
    \includegraphics[width=0.37\textwidth]{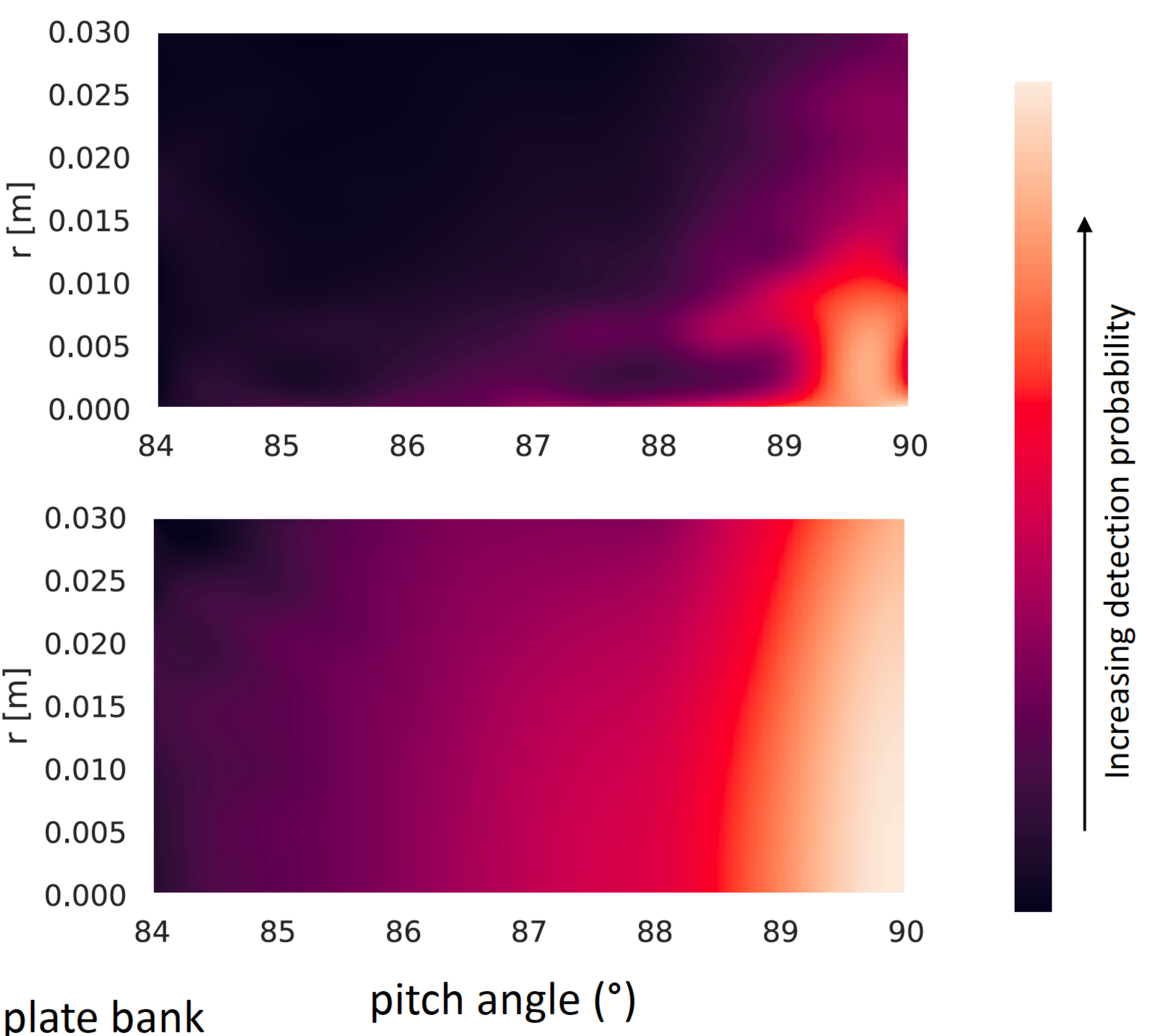}
    \caption{Left: Sideband structure for an electron in the FSCD detector. The sideband frequency structure reduces the available SNR to detect events in high noise environments.  Right: The radial and pitch angle efficiency reconstruction using DBF (top) and a template matched filter (bottom) for a single ring FSCD design.  Matched filtering is applied to the time-series data from all antennas.}
    \label{fig:sidebands}
  \end{figure}

In addition to trap optimization, advanced reconstruction techniques can take advantage of the sideband structure of these events to increase the sensitivity to events with low SNR.  Such techniques include artificial intelligence~\cite{Esfahani_2020} and matched filtering~\cite{steven2003fundamentals}.  Matched filtering is especially promising for reconstructing events, provided a sufficient and accurate template library is available.  An example comparison on the performance for reconstructing events with different pitch angles and radii can be seen in Fig.~\ref{fig:sidebands}.

\subsection{Cavity Resonator Design}
\label{sec:Cavity}

The cavity approach can potentially provide high signal-to-noise performance with a single (or few) channel readout.  As the volume and frequency are completely coupled in the cavity scheme, the cavity greatly benefits from operating at much lower frequencies/magnetic field strengths.  A detailed study of the CRES technique at various frequencies has shown a better expected performance at lower frequencies, especially for an atomic T experiment. 

\begin{figure}[htb]
   \centering
   \includegraphics[width=6in]{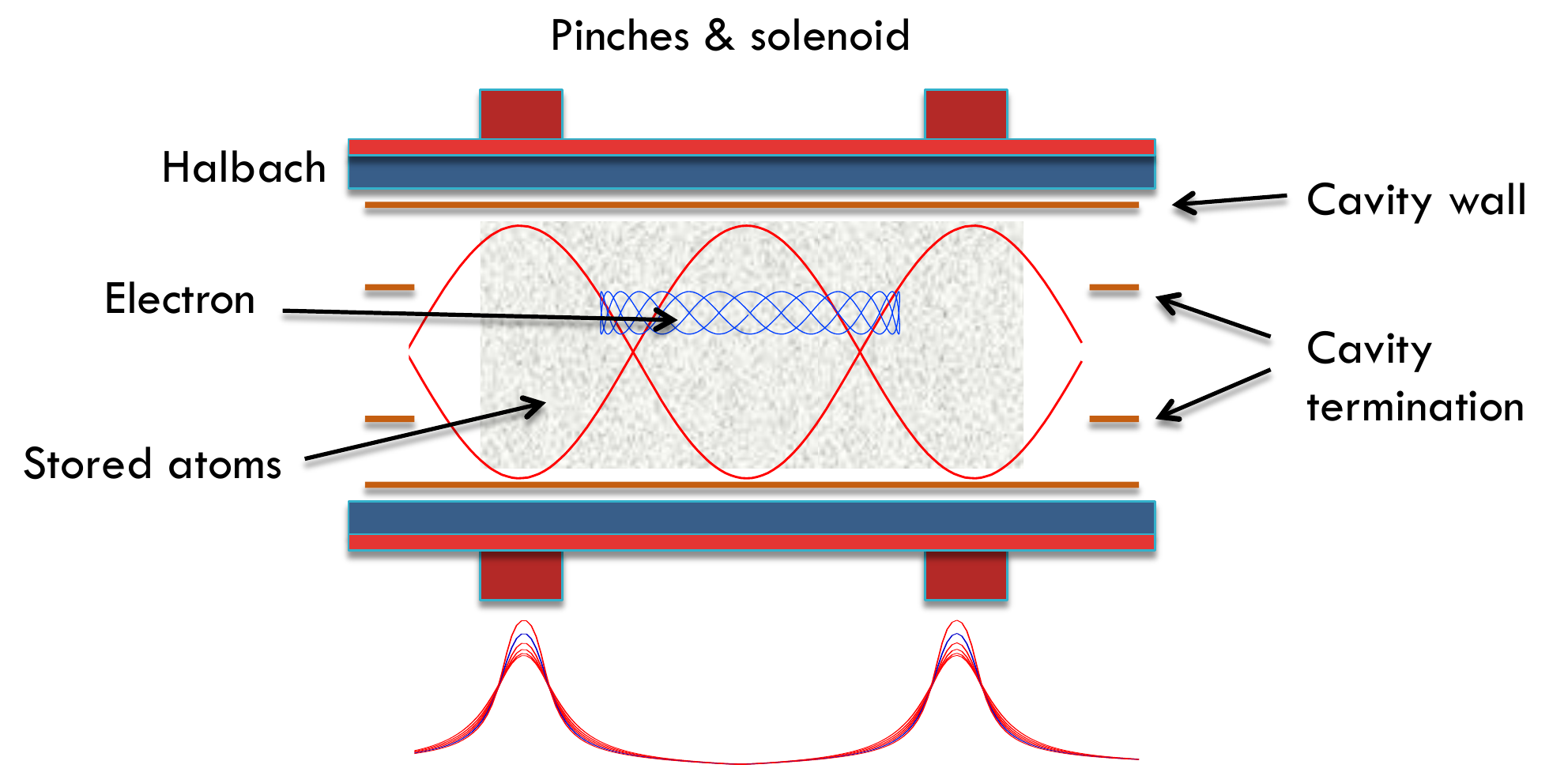}
   \caption{%Basic scheme of a resonant cavity readout for Project 8.
   Top: Basic scheme of a terminated, open-ended resonant cavity for Project~8. Bottom: Illustration of the trapping field created by the pinches and solenoid.}
   \label{fig:cavitycartoon}
\end{figure}

The elements of a CRES design based on a microwave resonant cavity are shown in Fig.~\ref{fig:cavitycartoon}.

In the following, we enumerate points in the design philosophy:

\begin{itemize}
    \item An electron in cyclotron and axial motion between the pinch coils can excite a standing-wave resonant mode in the cavity.  The Doppler effect is well tempered in a cavity because the phase velocity is very large and the modulation index $h\simeq 1$ for all pitch angles in the lowest axial mode.  A large range of pitch angles can therefore be detected, a necessity for high efficiency.
    
    \item A low frequency of operation is envisaged, 1 GHz or less, because atomic trap losses are dominated by dipole-dipole interactions. Those losses scale approximately as the square root of the magnetic field in the range 0.01 to 1 T \cite{Lagendijk_PhysRevB.33.626} and linearly with the atomic density.  Moreover, the cavity volume scales as the wavelength cubed, leading to a net scaling of source strength at least as the wavelength squared, if the usable density scales inversely as the wavelength.
    
    \item Cavities are usually closed containers, which is not a practical solution for Project 8.  The ends need to be open to permit pumping, beam injection, calibration, and other access needs.  Open-ended cavities are known and have been used in specialized applications \cite{thorn1959,freethey1959}. 
    Radiation from the ends is limited and Q is raised by means of suitable obstructions. A Q as high as 8000 was reported by \cite{freethey1959} in a cavity with a closed Q of 11000 with 92\% of the endplate material judiciously removed.  
    
    \item A large trap volume is needed for a sensitive experiment.  Even at the lowest practical frequencies, the desired trap dimensions imply a multimode cavity.  To avoid a complicated mode structure, we turn to the use of a ``mode-filtered'' cavity developed by NIST that consists of a wire winding or a stack of insulated rings \cite{NISThelicalcavity1993}.  This construction supports only modes of the form TE$_{0np}$ for which the wall currents are circumferential, suppressing all others, and giving a simple mode structure in a large cavity.
    \item As illustrated in Fig.~\ref{fig:cavitycartoon}, the electron should couple appreciably to a standing wave only in a single antinode.  Its axial motion modulates the coupling at twice the axial frequency, leading to AM sidebands.  The other two antinodes are in locations where the magnetic field is higher and the cyclotron frequencies do not excite the mode. The volume efficiency is optimized with a cavity having a single antinode along the axis, rather than the three as shown, if the magnetic design allows it.  
\end{itemize}
    
\subsubsection*{Mode-filtered cavities}

Solution of the field equations in a cylindrical cavity of radius $R$ and length $L$ leads to resonant modes that are either transverse electric (TE) or transverse magnetic (TM). A cylindrical cavity that is made in the form of a helical winding of insulated wire or a stack of insulated rings can only support modes for which the side-wall current is circumferential \cite{NISThelicalcavity1993}.  %All modes with longitudinal currents are strongly suppressed (by 30 db or so). 
The only modes with purely circumferential wall currents are TE$_{0n,p}$.  All magnetic modes, and other TE modes such as TE$_{11,p}$, have longitudinal currents and are strongly suppressed. The resonant frequencies of the TE modes, to which electrons couple well, are:
\begin{eqnarray}
f_{mnp}^2 (\rm TE)&=&\frac{c^2}{4\pi^2}\left[\left(\frac{X^\prime_{mn}}{R}\right)^2+\left(\frac{p\pi}{L}\right)^2\right] \label{eq:TE} \\
m&=&0,1,2,...;\ n=1,2,3,...;\ p=1,2,3,... \nonumber
\end{eqnarray}
The quantity $X^\prime_{mn}$ is the $n$th zero of the derivative of the $m$th Bessel function. The index $p$ counts the number of antinodes along the $z$ axis. 
Expressing the frequency in terms of the free-space wavelength $\lambda_0$, the physical volume of the cavity is
\begin{eqnarray}
V &=& \frac{(X^\prime_{mn})^2L^3}{\pi}\left[4\left(\frac{L}{\lambda_0}\right)^2-p^2\right]^{-1}.
\end{eqnarray}
Since we are free to choose $L$ and integer values of $p$, the volume can be made infinitely large, even for low mode indices $mnp$.  However, the mode density also becomes infinitely high because (from Eq.~\ref{eq:TE}) other modes with different $mnp$ produce resonances at neighboring frequencies.    Nevertheless, quite long cavities can be used.  This allows for larger volumes in the form of long cavities up to the point where the mode density becomes prohibitive.

The full span of the tritium spectrum in frequency is fractionally $(\gamma-1)/\gamma$, about 3.5\%.  This provides a benchmark for how closely intruder modes and neighboring p modes can be positioned relative to the mode being used for spectroscopy. Positioning the endpoint at the frequency of the $p=1$ mode and  zero kinetic energy at $p=2$ would prevent the intense spectrum of tritium from overwhelming the receiver.  As described below, the readout loop can be placed so as to reject the even-order modes, which means the zero-energy mode could be $p=3$ instead of $p=2$, doubling the available volume.  Even closer spacing could be used if the high-rate signals from deep in the tritium spectrum (but still not at the frequency used for endpoint studies) can be tolerated.    For an atomic tritium experiment at 1 GHz,  Table~\ref{tab:cavitydimensions} lists two examples of the cavity dimensions and the frequencies of other modes that lie closest to the selected one ($p=1$).  The choice of length is based on the desired mode separation: shorter spaces the modes out more but reduces the volume. 

\begin{table}[htb]
    \centering
    \caption[width=0.7\textwidth]{Dimensions of 1-GHz cavity, and frequencies of neighboring modes.    In the first section, the separation chosen is $\Delta f=f_{mnp}/72$.  Only 1 mode beside TE$_{01,1}$,  namely  TE$_{01,2}$, falls within the span of the tritium spectrum.  In the second section, the mode separation chosen is $\Delta f=f_{mnp}/266.5$ simply to produce a desired 3-m length. }
    \vspace{0.1in}
    \begin{tabular}{lcccc}
    \hline\hline
    Length, m & 1.5588 \\
    Radius, m & 0.1838 \\
    Volume, m$^3$ & 0.1654   \\
    Mode: & TE$_{01,1}$ & TE$_{01,2}$ & TE$_{01,3}$ & TE$_{02,1}$ \\
    Frequency, kHz: & 1000000 & 1013794 & 1036375 & 1824979 \\
    \hline
    Length, m & 2.9991 \\
    Radius, m & 0.1832 \\
    Volume, m$^3$ & 0.3161   \\
    Mode: & TE$_{01,1}$ & TE$_{01,2}$ & TE$_{01,3}$ & TE$_{02,1}$ \\
    Frequency, kHz: & 1000000 & 1003745 & 1009957 & 1829329 \\
        \hline\hline
    \end{tabular}
    \label{tab:cavitydimensions}
\end{table}

\subsubsection*{Doppler effect in cavities}

% In a waveguide or antenna detection scheme, the axial motion of a trapped electron results in frequency modulation and Doppler sidebands. The sideband structure is characterized by the modulation index $h$, which is the shift in frequency $\Delta f$ divided by the frequency of the shift, $f_a$, which in the present case is the axial frequency.  The shift in frequency is the Doppler shift, which in turn is the ratio of the electron's axial velocity $v_z$ to the phase velocity $v_p$ in the guide.  The sideband structure is very simple for $h\le 1$, a carrier and two symmetrical sidebands spaced on either side by the axial frequency. The sideband power rises from zero at $h=0$ to 0.19 in each of the $\pm f_a$ sidebands and about 1\% in higher-order sidebands at $h=1$.  For higher modulation indicies the sideband power increases, higher-order sidebands grow, and the carrier power drops.  The carrier disappears entirely at $h=2.405$, the first zero in the zero-order Bessel function.
The cavity excitation is driven by the cyclotron motion of the electron generating travelling waves that are trapped to form a standing wave.  The generating current is Doppler shifted by the ratio of the axial velocity $v_z$ to the phase velocity $v_p$.  Both positive and negative shifts are present.  
%Applying these principles to the cavity, we first express Eq.~\ref{eq:TE} in terms of the free-space wavelength $\lambda$:
%\begin{eqnarray}
%\frac{1}{\lambda^2}&=&\frac{1}{4\pi^2}\left[\left(\frac{X^\prime_{mn}}{R}\right)^2+\left(\frac{p\pi}{L}\right)^2\right].
%\end{eqnarray}
%The cutoff wavelength $\lambda_c$ and the phase velocity $v_p$ in the corresponding waveguide are \cite{Terman55}:
%\begin{eqnarray}
%\lambda_c&=& \frac{2\pi R}{X^\prime_{mn}} \\
%\frac{v_p}{c}&=&\left[1-\left(\frac{\lambda}{\lambda_c}\right)^2\right]^{-1/2}. 
%\end{eqnarray}
%Hence,
%\begin{eqnarray}
%\frac{1}{\lambda^2}&=&\frac{1}{\lambda_c^2}+\frac{p^2}{4L^2}.
%\end{eqnarray}
%In an ideal box trap of extent $L$, the axial frequency $f_a=v_z/2L$ and the Doppler shift is $\Delta f_c/f_c= \pm v_z/v_p$.  Gathering these, 
One finds that the modulation index becomes
%\begin{eqnarray}
$h = p.$
%\end{eqnarray}
In a cavity operating in the lowest axial mode the modulation index is therefore $\simeq 1$, independent of the frequency or pitch angle.  

This is an ideal outcome, with a pair of sidebands each having 1/3 the power of the carrier and very little power in higher-order sidebands.  Detectable sidebands are essential in the process of deriving the energy of an electron from its cyclotron- and axial-frequency spectrum, while excessive frequency modulation would make the spectrum too complex.  The sideband power will be modified  by amplitude modulation induced by the coupling of the electron in its axial motion to the cavity's axial mode structure, a half sine wave. More detailed studies are underway to fully understand the electron coupling into the available cavity modes, but it is expected that pitch angle and energy reconstruction should be possible over a wide range.  Selection of the $p=1$ mode is favored both to limit the modulation index and to maximize the volume efficiency.

\subsubsection*{Volume acceptance}

\begin{figure}[htb]
   \centering
   \includegraphics[width=3.1in]{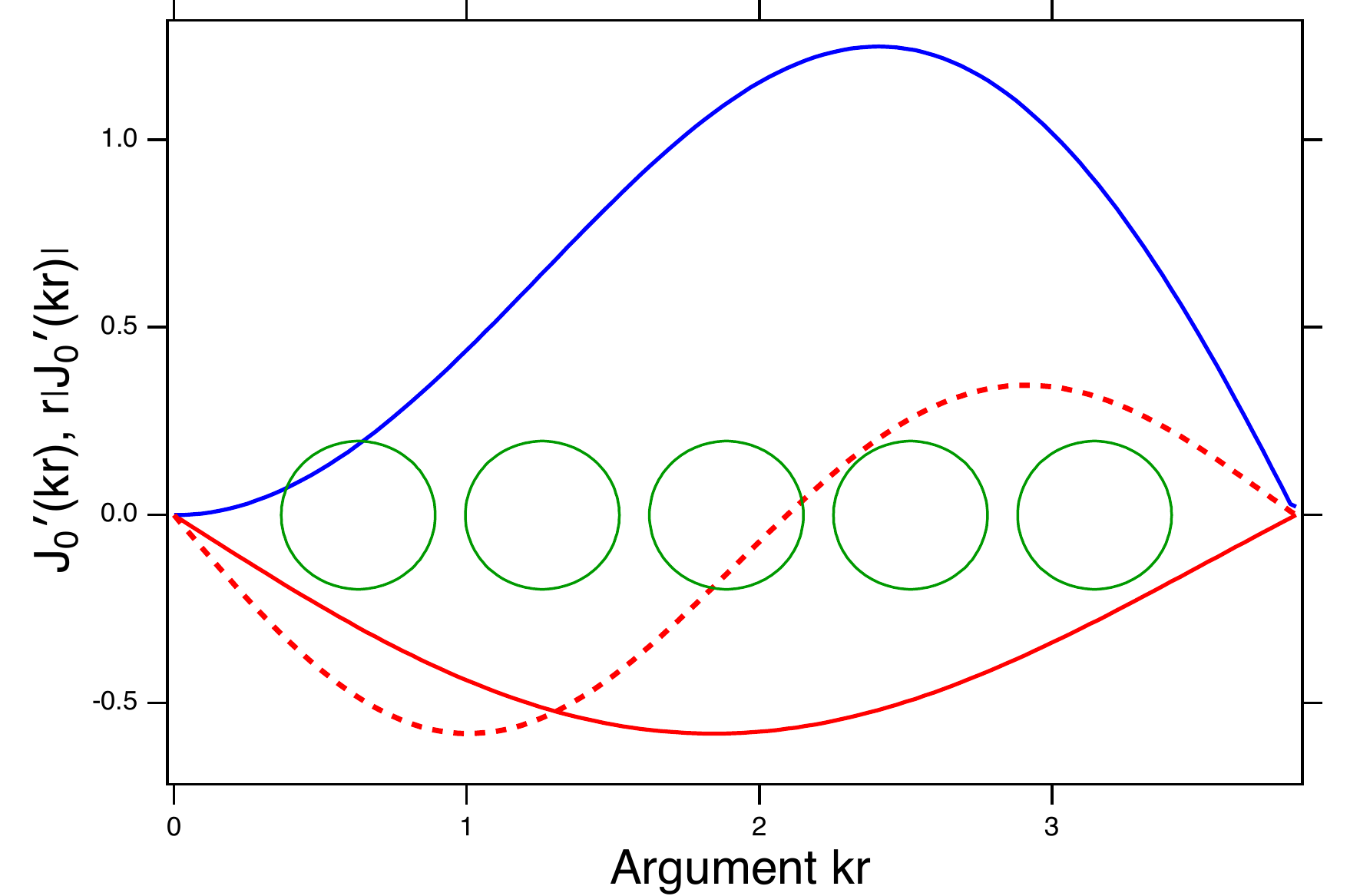}
   \includegraphics[width=3in]{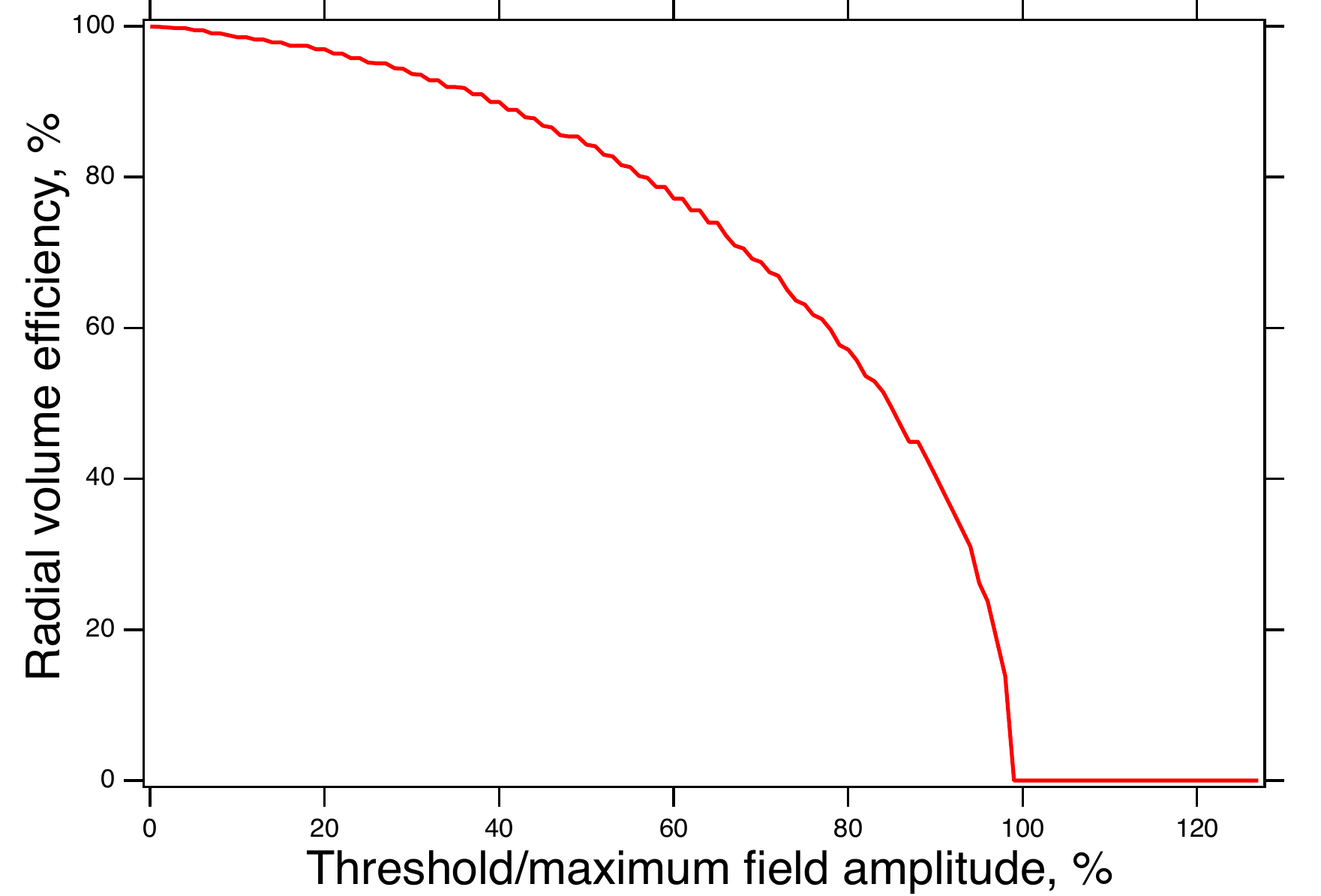}
   \caption{Top: (red) Derivative of $J_{0}(kr)$; (blue) volume-weighted field strength; (dotted red) second-harmonic excitation of TE$_{02,16}$; (green circles) Larmor-radius orbits. Bottom: Radial volume efficiency as a function of threshold.}
   \label{fig:besselplots}
\end{figure}

The detection efficiency is the ratio of the amount of source gas that produces detectable events at the endpoint to the total amount of trapped source gas.  It has three main components,  volume acceptance, pitch-angle acceptance, and track and event reconstruction efficiency.  The TE$_{01,p}$ modes have a radial distribution of the electric field strength that is zero on the axis and at the wall, with a maximum at the peak of the derivative of the Bessel function $J_{0}(kr)$, where $k$ is a constant determined from the boundary conditions for the mode. The derivative of the Bessel function is given by
\begin{eqnarray}
\frac{d}{dx}J_0(x)&=&\frac{1}{2}[J_{-1}(x)-J_1(x)].
\end{eqnarray}
% The distribution is well suited to the CRES atomic trap because electrons on the axis are not trapped in any case, since atoms must be injected without reflection.  The pinch field at one end should be zero on axis to allow for beam injection.  The field near the wall is quite non-uniform because of the atomic-trap magnet, and electrons in that region are also not useful.  
Figure~\ref{fig:besselplots} shows the electric field profile as a function of radius, the volume-weighted profile, and the radial volume efficiency as a function of threshold.
  If the threshold is set at 70.7\%, the radial volume efficiency is 67\% (this figure does not include the power dependence from pitch angle). Also shown are cyclotron orbits:  the Larmor radius $r_c$ bears a fixed ratio to the cavity radius $R$ equal to $\beta/X^\prime_{01}$ (in the limit $p/L\rightarrow0$).  Because the Larmor radius is non-negligible compared to the cavity radius, appreciable excitation of the second harmonic is possible, proportional to $J^{\prime\prime}_0 J^{\prime}_0$.  In a  cylindrical cavity in TE$_{01,1}$ with the dimensions from Table~\ref{tab:cavitydimensions} (bottom), the second harmonic is between TE$_{02,16}$ and TE$_{02,17}$.  To avoid absorption dips in the spectrum, it is important to move the second harmonic resonances away from the region of interest, which can be done by small changes in length or the shape of the ends. 

The axial volume efficiency depends on the location of the trap pinch coils.  In order that electrons from outside the trap are reflected, to insure a low background, the trap coils should be closer together than the cavity length, but not too close to avoid raising $h$ or degrading the volume efficiency. If trapped electrons thus turn around before reaching the cavity ends, at about the same place that atoms are reflected, the axial volume efficiency approaches 100\%.

\subsubsection*{Assembly}

\begin{figure}[b!]
   \centering
   \includegraphics[width=0.75\textwidth]{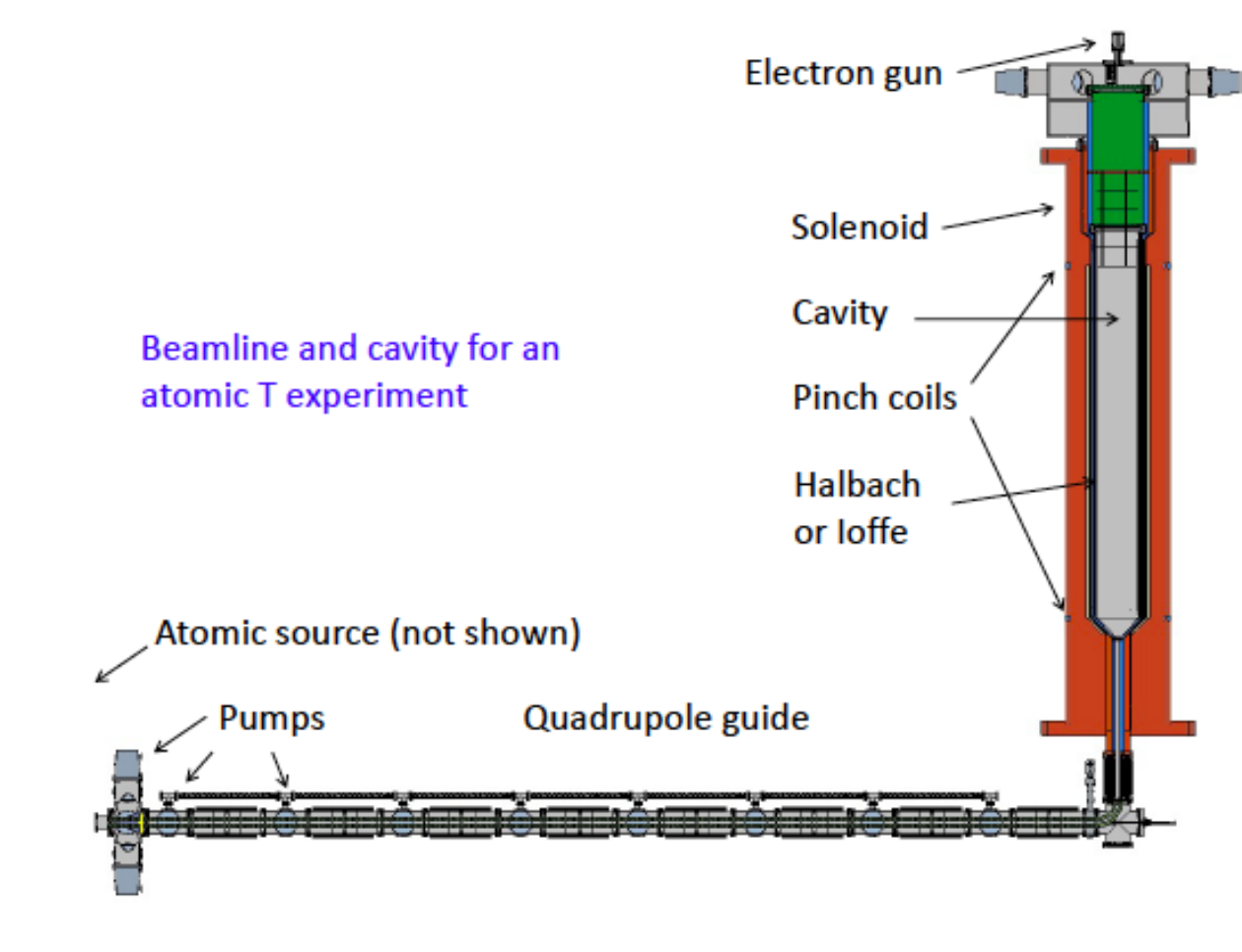}
    \caption{Cavity resonator design, shown with atomic beamline and magnetic field system.}
   \label{fig:cavity_beamline}
\end{figure}

The cavity concept culminates in a design as shown in Fig.~\ref{fig:cavity_beamline}. 
The cavity is shown for simplicity as a smooth  cylinder with a conical lower end and an upper end
connected to a cavity support tube made of thin-wall stainless steel.  The cavity is oriented vertically to allow magnetogravitational trapping of cold atomic tritium.  The atomic trapping field would be provided either by a Halbach array (permanent magnets) or Ioffe-Pritchard coil (superconducting).  The current trapping magnet design is under study.  An electron calibration source (electron gun) would feed from the top of the cavity to provide monoenergetic electrons into the main volume for Mott scattering into trapped orbits from a background gas such as He.  The main CRES field would be provided by a simple solenoid, augmented by a pair of larger-diameter field-shaping coils at the ends and a pair of pinch coils near the cavity ends.  The field strength at the center would reach 0.04 T (1 GHz).

\section{Atomic Tritium Source}\label{sec:atomic}

{\em Most tritium beta decay experiments to date have used molecular tritium as their primary target.  Although comparatively simpler to use in contrast to their atomic counterparts, molecular sources introduce an irreducible uncertainty in the beta decay spectrum that limits the sensitivity of such experiments to above 100 meV.  In order to reach the target sensitivity of $m_\beta \ge 40$ meV/c$^2$, a switch to a high purity atomic source is necessary.  The Project 8 collaboration is designing an atomic tritium source that is magnetogravitationally confined within the detection vessel in order to overcome this limitation.}

\begin{figure}[b!]
    \centering
    \includegraphics[width=0.53\textwidth]{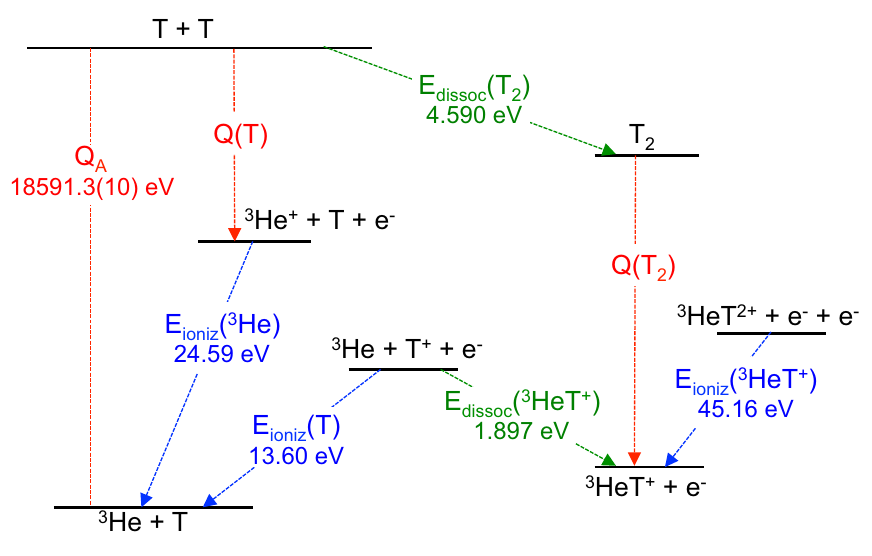}
    \hspace{3em}\includegraphics[width=0.35\textwidth]{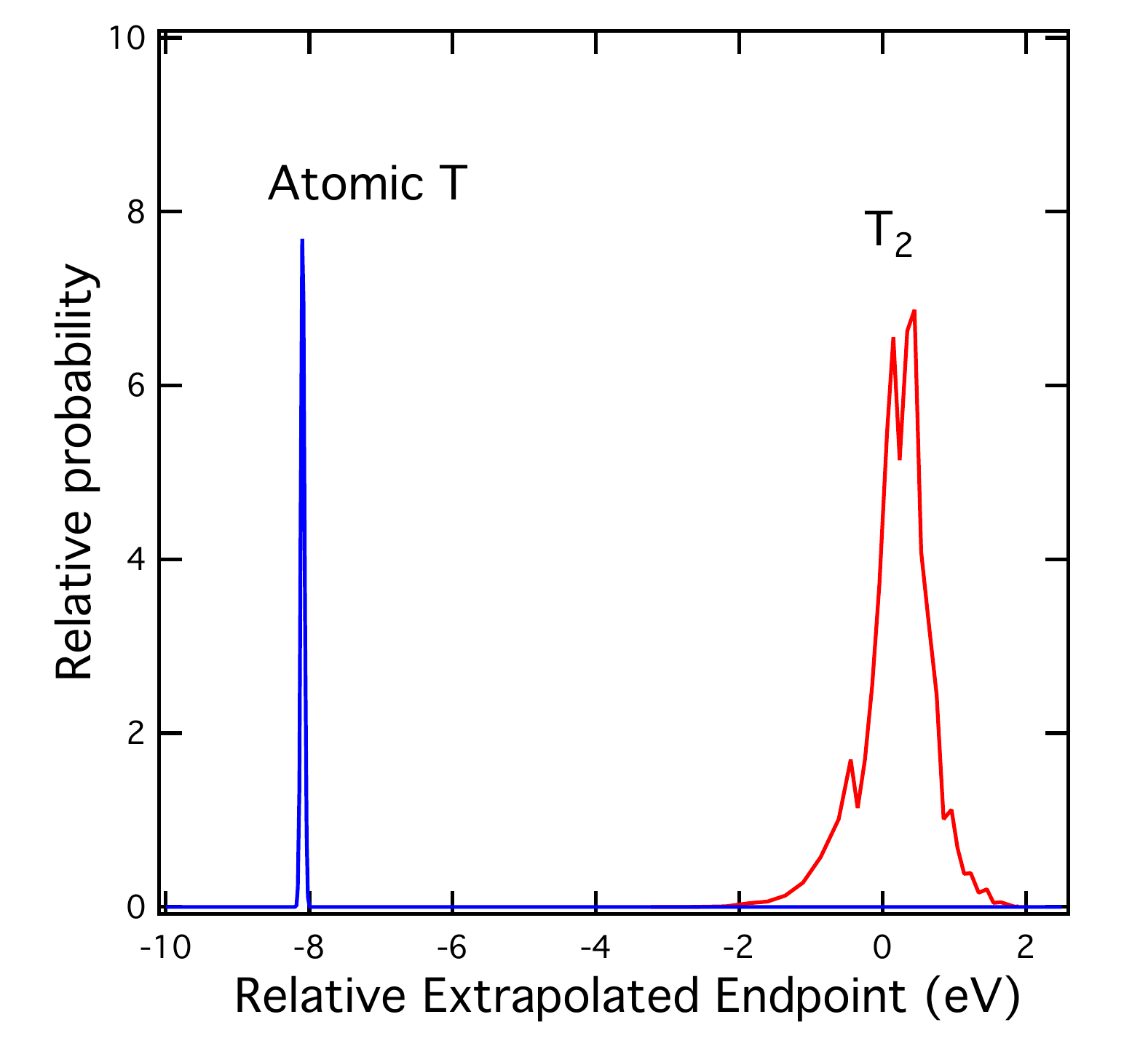}
    \caption{Left: Energy levels of the formation and decay of the T$_2$ molecule. Figure from Bodine et al.\cite{Bodine:2015aa}. Right: Theoretical comparison of atomic and molecular final-state distributions~\cite{Saenz:2000fk}.}
    \label{fig:3HeTDiagram}
\end{figure}

As discussed above in Section~\ref{sec:beta_decay}, uncertainty in tritium beta decay final states can introduce a systematic error that is indistinguishable from the contribution of neutrino mass~\cite{Robertson:1988aa}. Unphysical negative values for $m_\beta^2$ determined by some past experiments are now confirmed to have been caused by insufficient knowledge of the final states available at that time, as had been suspected~\cite{Bodine:2015aa}.  If a molecular source is used, outgoing beta-decay electrons share energy with multiple vibrational and rotational states of the resulting \ce{^3HeT^+} molecule, shown in Figure~\ref{fig:3HeTDiagram}. 

The tritium endpoint method is very sensitive to uncertainty in the final state distribution. For example, KATRIN relies on 1\% precision in its variance in order to meet its goals.  The smearing effect of final states also exacts a statistical penalty.  Because of the theoretical uncertainties and experimental systematic effects associated with molecular tritium, any experiment that aspires to improve much upon KATRIN's targeted \SI{0.2}{\electronvolt} sensitivity will have to use an atomic tritium source.

The molecular tritium endpoint is \SI{10}{\electronvolt} higher than that for atoms. Because the differential spectrum rises like $\epsilon^2$ for $\epsilon \gtrsim m_{\beta}$ (see Equation~\ref{eqn:tritiumBetaDecay}), a small contamination of molecules would create a large background to the atomic tritium spectrum.
Project 8 will therefore have a very stringent upper limit on the ratio of densities $n({\rm T}_2)/n({\rm T}) \lesssim 10^{-4}$ to preserve the benefits of CRES's otherwise low-background nature.
Even once the atomic tritium is produced, it must be trapped in a way that reduces the rate of recombination.

Recombination occurs in three distinct processes that scale as the first three powers of gas density, $n({\rm T})$~\cite{Walraven:1982aa}. The first two, proportional to $n$ and $n^2$, occur on surfaces. The third, proportional to $n^3$, occurs in the gas volume. At the vacuum pressures below \SI{e-5}{\milli\bar} required by CRES, volume recombination is negligible.  Conversely, recombination on material vessel walls occurs so rapidly that only magnetic and gravitational potentials can be used for confinement. Project 8 will use a magnetogravitational trap, which confines T but not \ce{T2}.

The tritium atoms are created by thermal dissociation of \ce{T2} molecules. They are emitted from the so-called ``thermal cracker’' at 2300-\SI{2600}{\kelvin}, and are cooled to $\sim$\SI{1}{\milli\kelvin} in three stages. In the first cooling stage, a series of thermally isolated “accommodator” tubes held at successively lower temperatures cool atoms and molecules by interactions with the walls.  Accommodator cooling can reduce the temperature to 10 to \SI{40}{\kelvin}, limited by the need to maintain atoms and molecules in the gas phase.  The accommodator is the last site where wall interactions can be tolerated.

\begin{figure}
    \centering
    \includegraphics[width=0.6\textwidth]{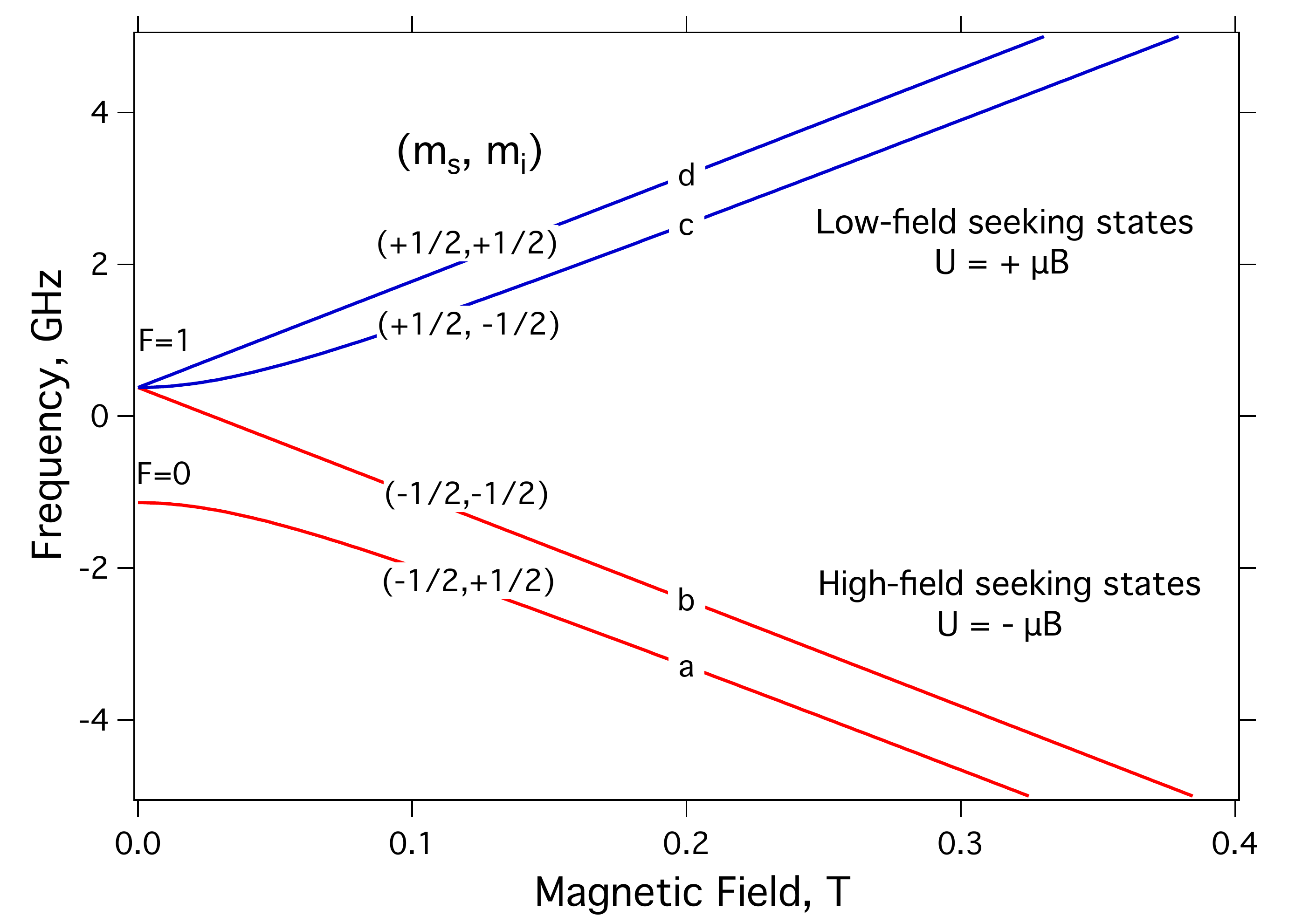}
    \caption{Energy levels of the hyperfine states of atomic tritium as a function of magnetic field.}
    \label{fig:FieldSeekingStates}
\end{figure}

Subsequent cooling and trapping requires that the atoms be magnetically entrained.    As shown in Figure~\ref{fig:FieldSeekingStates}, atoms in the `c' and `d' hyperfine states (``low-field-seeking states'')  gain potential energy with increasing magnetic field strength and can be confined in a quadrupole or higher-order radial field.  Axial confinement can be provided by `pinch' coils or by higher-order multipoles.  Both atoms and electrons are confined in such arrangements but at the low magnetic fields optimal for CRES, the trapping requirements are very different.  Electron trapping requires only axial field barriers, and the pitch-angle range trapped depends only on the ratio of that pinch field to the center field.  Atom trapping, on the other hand, requires high absolute field strengths at the `walls' depending on the temperature of the gas.    Molecules entering or formed by recombination in the volume of the trap will promptly escape, since their magnetic moments are orders of magnitude smaller, and can be cryopumped by the inner surface walls. Betas emitted from the walls will not contribute to the CRES signal because they undergo, at most, one cyclotron orbit before impacting the wall.
% Next, a velocity and state selector (VSS) uses magnetic quadrupole guides and/or a thin magnetic lens, such that any species without the correct combination of mass, magnetic moment, and velocity will be filtered by the guide or lens.

In principle, the cold atoms can be obtained by taking a momentum-analyzed slice of the Maxwell-Boltzmann distribution from the accommodator.    
Atoms at 10 to \SI{40}{\kelvin} are still far too hot to be well matched to the magnetic trap at  $\sim$\SI{1}{\milli\kelvin}, and the efficiency is very low.  The same problem was encountered in the 1980s and 1990s in the experiments to make a Bose-Einstein condensate of hydrogen atoms.  Success was eventually achieved with a multistep cooling strategy that was set forth by Hess \cite{Hess_PhysRevB.34.3476}.  

The techniques used for hydrogen cooling, however, do not easily lend themselves to work with tritium.  For example, tritium atoms cannot be cooled on superfluid helium because the surface adsorption energy is too high \cite{SilveraWalraven1986}.  Nor can fluorocarbon accommodators be used because of attack by tritium radioactivity. However, the  evaporative cooling step described by Hess can nevertheless be envisioned, not in the static setup used by the BEC researchers, but in a dynamic beamline environment. 
We consider a quadrupole beamline fed by atoms from an accommodator.  The atoms are cooled by evaporation over the magnetic wall as they move along the beamline.  Atomic beams have been cooled by forced evaporation induced by RF fields, and used for the production of cold beams of alkali metal atoms \cite{evapcoolChaudhuriJPCS2007,evapcoolMandonnetEPJ2000,evapcoolLahayePhysRevA.72.033411}.  In our experiment, the atomic density is high enough to cool the atoms by reduction of the trapping potential along the beam line. Cooling the longitudinal motion requires many quasi-random magnetic perturbations to convert the longitudinal into transverse motion, or the use of `Sisyphus cooling', a sequence of magnetic hills with spin flips induced at the top and bottom.  A different, more compact, geometry would be a superconducting quadrupole in a spiral form.  This evaporative scheme is currently under investigation, and early studies suggest that sufficient cooling capacity may be attainable.

Finally, a magnetic or magnetogravitational trap will hold and store the atomic tritium, while minimizing contact with the vessel surface where recombination is likely.  The magnetic radial trap will be either a superconducting Ioffe configuration, or a Halbach array of permanent magnets. A Ioffe trap consists of counter-propagating axial currents on the surface of a cylinder. A Halbach array consists of segments of permanent magnets with a periodic pole structure. In both cases the resulting field is strong close to the physical surface but drops off quickly towards the center. If placed inside a solenoid, the result is a highly uniform $B$ field required by CRES in the majority of the volume, with strong gradients at the walls to repel spin-polarized atoms.  A magnetic wall height of  \SI{50}{\milli\tesla} will trap atoms below \SI{1}{\milli\kelvin}.
\section{Sensitivity}\label{sec:Sensitivity}

{\em Project 8 aims to measure the absolute neutrino mass scale, down to a final sensitivity of $m_\beta \le 40$ meV/c$^2$ at 90\% confidence level.  Sensitivity studies have shown that such a sensitivity is achievable provided that one makes use of an atomic tritium source and the experiment reaches sufficiently high activity and resolution.  At such a sensitivity level, the experiment can also provide important information on the existence of eV-scale sterile neutrinos and the neutrino mass ordering.}

\subsection{Sensitivity Estimation}\label{subsec:Sensitivity}

The main drivers for the ultimate sensitivity of a neutrino mass experiment are the following four factors: (a) statistical accuracy, (b) energy resolution, (c) background, and (d) systematic uncertainties. The CRES technique gains much of its statistical power from its function as a {\em differential spectrometer}, i.e., the spectrum is counted as a whole, with each event sorted by energy (microcalorimeters are also of this type). 

The statistical sensitivity to the neutrino mass is fundamentally determined by the number of events in a small `analysis window' $\Delta E$ near the endpoint, where the neutrino mass has the most significant impact. The count rate near the endpoint is given by Eq.~(\ref{eqn:tritiumBetaDecay}), reproduced here:
\begin{equation}
\frac{dN}{d\epsilon} \approx 3rt \epsilon^2 \left( 1 - \frac{m_{\beta}^2}{2 \epsilon^2} \right).
\end{equation}
Integrating across $\Delta E$ provides an estimate of the number of signal events in the region. Any differential backgrounds within $\Delta E$ can also be included:
\begin{eqnarray}
N_{\rm tot} &=& rt (\Delta E)^3\left[1-\frac{3}{2}\frac{m_\beta^2}{(\Delta E)^2}\right] +bt\Delta E.
\end{eqnarray}
The statistical uncertainty $\sigma_{m_\beta^2} $ is then related to the variance in the total number of events:
\begin{eqnarray}
\sigma_{m_\beta^2} &\simeq & \frac{2}{3rt}\sqrt{rt\Delta E + \frac{bt}{\Delta E}}.  \label{eq:sig}
\end{eqnarray}
There is also an optimum choice of $\Delta E$ that minimizes the statistical uncertainty,
\begin{eqnarray}
\Delta E_{\rm opt} &=& \sqrt{\frac{b}{r}}.
\end{eqnarray}
For this choice,
\begin{equation}
 \sigma_{m_\beta^2} \equiv \sigma_{\rm opt} \simeq \frac{2^{3/2}b^{1/4}}{3t^{1/2}r^{3/4}}, 
\end{equation}

For a full derivation of this result, see Ref.~\cite{direct}.  As a practical matter, the ratio $b/r$ may be very small when rates are high or backgrounds low.  The optimum analysis window $\Delta E$ is then determined by other factors, such as the instrumental broadening with standard deviation $\sigma_{\rm instr}$---because the neutrino mass's effect on the spectrum is now smeared over this larger interval.  In turn, improving the instrumental resolution beyond a certain point is not useful if one encounters a limit set by  final-state distribution (FSD) broadening.  In the decay of T$_2$ to T-$^3$He$^+$, the molecular final-state distribution of the ground-state rotational and vibrational manifold has a standard deviation $\sigma_{\rm FSD}\simeq 0.4$ eV~\cite{PhysRevLett.84.242,Bodine:2015aa}.    The temperature also plays a role through translational Doppler broadening $\sigma_{\rm trans}$  \cite{Bodine:2015aa}. The quadrature sum of these contributions forms a basis for fixing $\Delta E$:
\begin{eqnarray}
\Delta E &=& \sqrt{ \frac{b}{r} + C^2(\sigma_{\rm FSD}^2+ \sigma_{\rm trans}^2 +\sigma_{\rm instr}^2 + ... )}  \label{eq:eqten}
\end{eqnarray}
where $C=\sqrt{8\ln{2}}=2.35$.  Many factors contribute to the instrumental resolution $\sigma_{\rm instr}$, such as magnetic field inhomogeneities, signal-to-noise ratio, electron-gas scattering uncertainties, and plasma effects.  All such resolution contributions have associated uncertainties that are indistinguishable from neutrino-mass effects and thus set a floor on the sensitivity even with the highest statistical precision \cite{direct}.

The observed rate $r$ in Eq.~\ref{eq:sig} increases with the effective volume $V_{\rm eff}$ of the source\footnote{Effective volume is defined as a volume filled with the source density from which every beta decay electron in the region of interest is detected.} and with the density $n$ of tritium gas.  An additional optimization for $n$ is therefore necessary. If $n$ is too high, particle collisions will reduce observation times, increasing the uncertainty on cyclotron-frequency measurement, and in turn degrading energy resolution and increasing its associated systematic error. The optimum density is found by minimizing an approximate formula for the uncertainty in $m_\beta^2$, which reveals the competing dependencies on $n$. 

As discussed previously, the replacement of molecular tritium with atomic tritium removes final-state uncertainties which would otherwise limit neutrino mass sensitivity to $\approx$ 100 meV.  However, any atomic experiment would need to ensure that the level of contamination from molecular tritium is kept at extremely low levels, since the endpoint energy for atomic tritium lies effectively about 8 eV {\em below} the molecular endpoint.  Traces of molecular tritium thus manifest as an energy-dependent background and the molecular-to-atomic contamination level must be kept below 0.01\%.  %Given the propensity of tritium molecules to recombine on surfaces, the atomic tritium source needs to be magnetically contained away from the vessel ways, while any molecular tritium accumulation on the vessel is cryo-pumped away from the main volume.

\begin{figure}[hbt]
    \centering
    \includegraphics[width=0.75\textwidth]{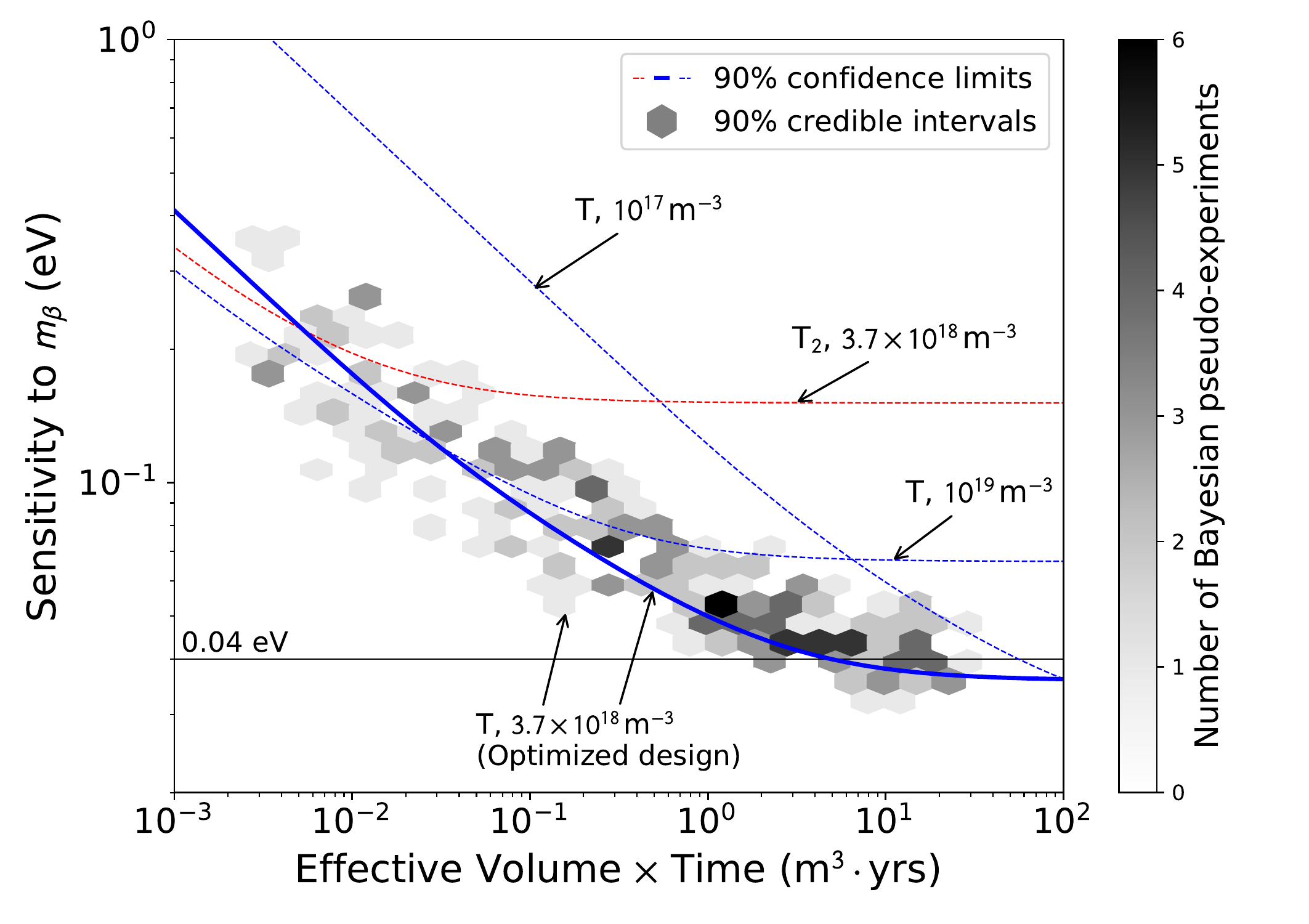}
    \caption{Dependence of mass sensitivity (width of 90\% confidence or credible intervals) on volume $\times$ efficiency $\times$ time. The plot assumes a scenario with instrumental resolution and uncertainty  $\sigma_{\rm instr} = 115 \pm 2$ meV.  Solid lines indicate  scenarios of different densities and gas compositions using an analytical approach, while the hexagonal dots are derived from a full Bayesian analysis. See Ref.~\cite{AshtariEsfahani:2020bfp} for additional details.}
    \label{fig:sensitivity}
\end{figure}

Although the above relations serve only as an approximation to the sensitivity, they track remarkably well with more comprehensive sensitivity analyses.  Project 8 has compared sensitivity projections from the analytical approach above with projections from detailed pseudo-data studies, using Bayesian techniques with a Markov-Chain Monte Carlo algorithm.  The two approaches agree well with each other (see Fig.~\ref{fig:sensitivity}).

\subsection{Projected Results}

\begin{figure}[b!]
    \centering
    \includegraphics[width=0.75\textwidth]{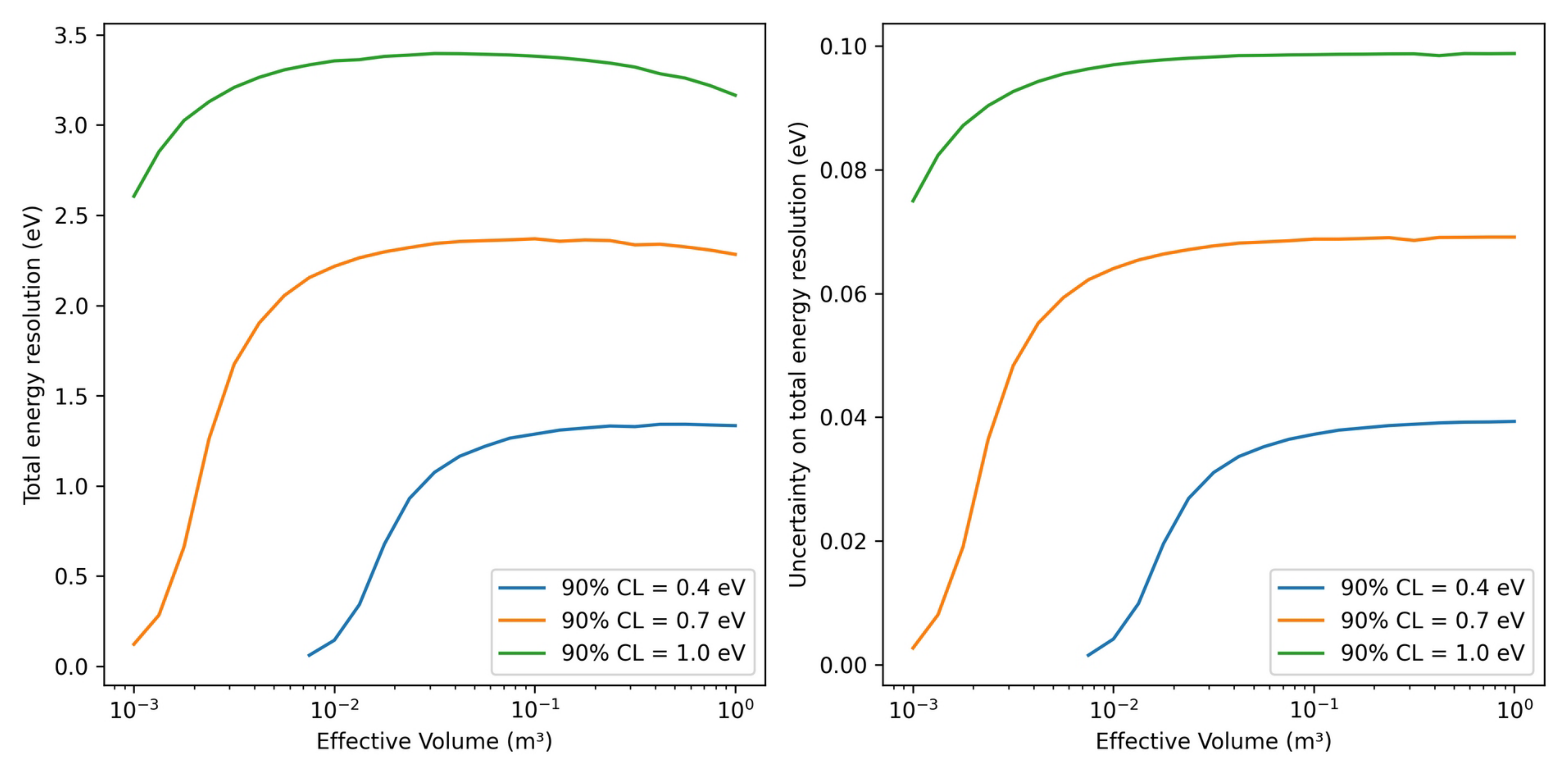} \\
    \includegraphics[width=0.75\textwidth]{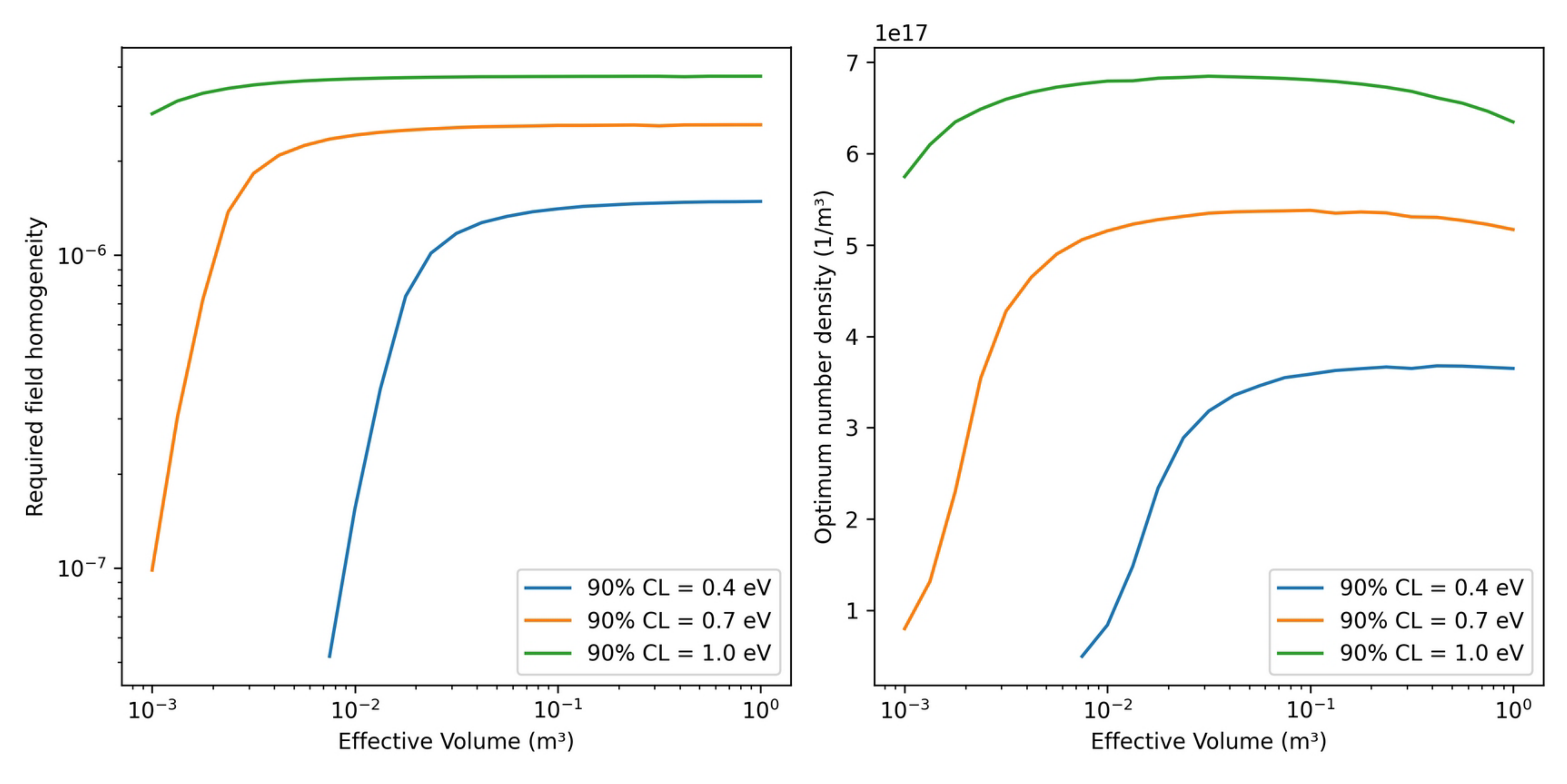}
    \caption{Curves of constant neutrino mass sensitivity, $m_\beta \le $ (1.0, 0.7, 0.4) eV/c$^2$ at 90\% CL for a 0.01-m$^3$ effective volume atomic tritium mass experiment with 1-year run time operated at 0.04 T (1 GHz) magnetic field with a noise temperature of 4 K.  Optimization curves are shown for the total energy resolution, the energy resolution error, magnetic field homogeneity, and density.}
    \label{fig:sensitivity_compare}
\end{figure}

\begin{figure}[b!]
    \centering
    \includegraphics[width=0.75\textwidth]{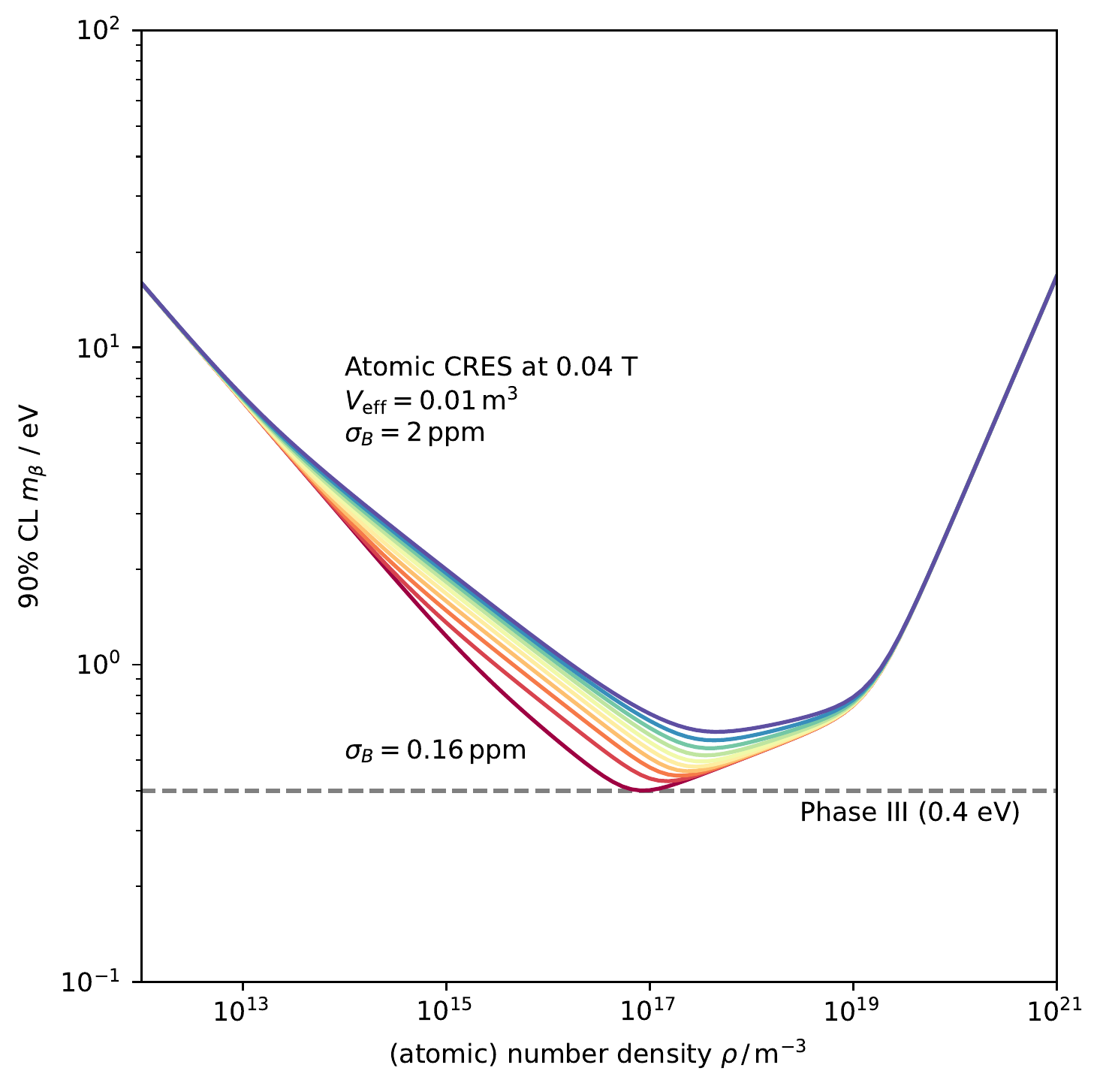}
    \caption{Dependence of mass sensitivity (90\% confidence limit) on density for a 1-GHz atomic tritium source with a 0.01-m$^3$ effective volume.  Different colored curves show a range of allowed field homogeneity resolutions under otherwise identical conditions.}\label{fig:sens_density}
\end{figure}

A number of interconnected factors ultimately determine the sensitivity of a CRES-style experiment.  For a given configuration (atom/molecule ratio, magnetic field, total volume, etc.), it is possible to determine the family of solutions which is consistent with a given neutrino mass sensitivity.  It is thus possible to determine the optimal density and energy resolution required in order to reach a target neutrino mass sensitivity. Fig.~\ref{fig:sensitivity_compare} 
illustrates this procedure for the case of the 1-GHz cylindrical cavity. Ranges of optimal density, magnetic field homogeneity, and total energy resolution are scanned as a function of effective volume. From such an optimization, it is possible to extract the projected neutrino mass sensitivity for a given experiment.  We make use of the Cramer-Rao Lower Bound (CRLB) to determine the relation between frequency/energy resolution and tritium density~\cite{bib:BuzinskyThesis,bib:CoverThomas}.  Results for a sample atomic tritium 1-GHz experiment are shown in Fig.~\ref{fig:sens_density}.

\begin{figure}[hbt]
    \centering
    \includegraphics[width=0.85\textwidth]{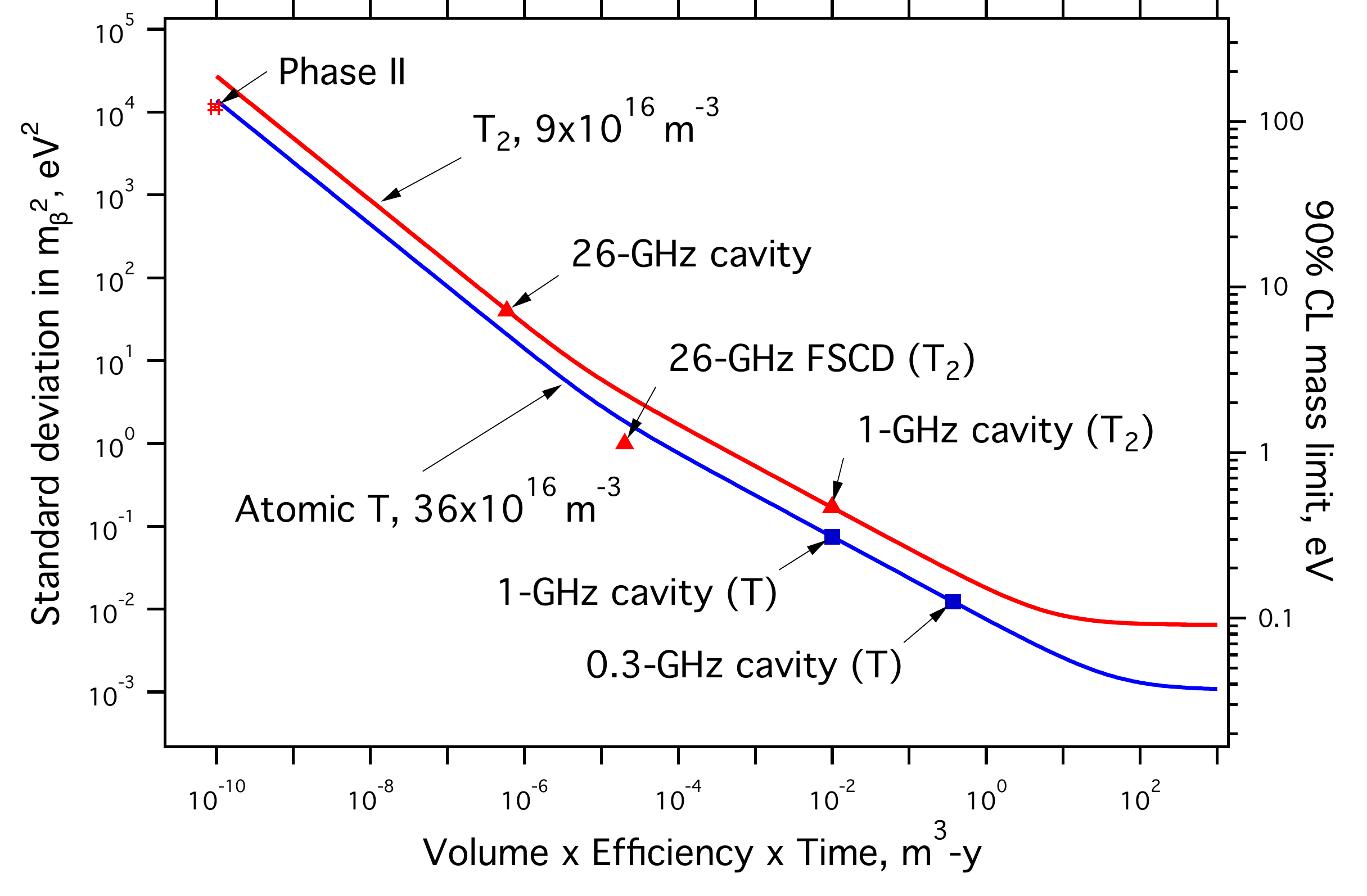}
    \caption{Dependence of mass sensitivity (90\% confidence-level upper limits) on volume $\times$ efficiency $\times$ time for  atomic and molecular tritium at the indicated densities.  Markers for the sensitivity measured in the Phase II waveguide experiment and for  demonstrator concepts are also shown.}\label{fig:sensitivity_p8}
\end{figure}

%The effective volume of available tritium decays determines the final sensitivity of Project 8. 
Fig.~\ref{fig:sensitivity_p8} shows the recent Phase II neutrino mass measurement together with the target sensitivity for future Project 8 demonstrators,
as a function of effective volume $\times$ time.  
%for several scenarios (atomic, molecular).
The Phase II measurement agrees fairly well with the analytic sensitivity prediction, but is about a factor of 2 more sensitive in $m_\beta^2$.  This may be due to the low statistics of Phase II because the analytic model assumes Gaussian statistics, or it may result from the analytic model making no use of the additional information on neutrino mass to be found in the spectrum shape just below the analysis window.  The \ce{T2} source density shown in the figure as the red curve is comparable to the density used in Phase II.  The blue curve is atomic T with the same stopping power.   

The specific choice of demonstrators is in progress. The 26-GHz cavity is not planned for use with tritium, and will be tested with an electron gun.  The 26-GHz FSCD antenna array is estimated for molecular tritium at a density 10 times higher than the curve shown in the figure, and thus falls below it.  It will be initially tested also with an electron gun, with a decision on tritium running deferred until then.   Larger demonstrators, such as the 1-GHz and 0.3-GHz cavities, can push the neutrino mass sensitivity into the regime presently accessible only to KATRIN with just one year of data taking with molecular and atomic tritium. Only one such demonstrator is envisaged, with the choice of operating frequency to be based on further analysis.  A demonstrator in that size range would have major physics objectives in its own right.  

Advances in signal processing using matched filters and machine learning are expected to make increased source densities usable.  An example is the 26-GHz FSCD demonstrator, which uses matched filters to work with higher density gas. These methods may be applicable to cavities as well.  

 To reach the inverted-ordering exclusion scale of $m_\beta \le 40$ meV/c$^2$ calls for a large-volume experiment.  Such an experiment may be realized by a very large free-space antenna array or multiple low-frequency cavities.  Large, monolithic systems are disfavored because they stretch the envelope of atomic-source capability and are more vulnerable to downtime.

In summary, the Project 8 collaboration is undertaking a staged approach to measuring neutrino mass down to the inverted hierarchy scale of 40 meV/c$^2$.  The CRES technique offers the statistical power of a differential spectrometer, high resolution, very low background, and the potential for introducing a new source with very low systematics, atomic tritium.  Achieving this goal will have significant impacts on the fields of nuclear physics, particle physics, and cosmology.

\section{Acknowledgments}\label{sec:ack}

This material is based upon work supported by the following sources: the U.S. Department of Energy Office of Science, Office of Nuclear Physics, under Award No.~DE-SC0020433 to Case Western Reserve University (CWRU), under Award No.~DE-SC0011091 to the Massachusetts Institute of Technology (MIT), under the Early Career Research Program to Pacific Northwest National Laboratory (PNNL), a multiprogram national laboratory operated by Battelle for the U.S. Department of Energy under Contract No.~DE-AC05-76RL01830, under Early Career Award No.~DE-SC0019088 to Pennsylvania State University, under Award No.~DE-FG02-97ER41020 to the University of Washington, and under Award No.~DE-SC0012654 to Yale University; the National Science Foundation under Award Nos.~PHY-1205100 to MIT; the Cluster of Excellence “Precision Physics, Fundamental Interactions, and Structure of Matter” (PRISMA+ EXC 2118/1) funded by the German Research Foundation (DFG) within the German Excellence Strategy (Project ID 39083149); the Laboratory Directed Research and Development (LDRD) 18-ERD-028 at Lawrence Livermore National Laboratory (LLNL), prepared by LLNL under Contract DE-AC52-07NA27344, LLNL-JRNL-817667; the LDRD program at PNNL; the University of Washington Royalty Research Foundation; Yale University; and the Karlsruhe Institute of Technology (KIT) Center Elementary Particle and Astroparticle Physics (KCETA). 

\bibliographystyle{unsrt_et_al}
\bibliography{papers}

\end{document}